\documentclass[aps,prx,groupedaddress,twocolumn, secnumarabic, amssymb,nobibnotes, floatfix, 10pt]{revtex4-2}
\pdfoutput=1

\usepackage[T1]{fontenc}
\usepackage[nolist]{acronym}
\usepackage{microtype}
\usepackage{amsmath}
\usepackage{bm}
\usepackage{siunitx}
\usepackage{color}
\usepackage[percent]{overpic}

\DeclareSIUnit{\ang}{\mbox{\normalfont\AA}}

\newcommand{\doh}{d_{\mathrm{OH}}}
\newcommand{\dohs}{d_{\mathrm{OHs}}}
\newcommand{\doha}{d_{\mathrm{OHa}}}
\newcommand{\hoh}{\phi_{\mathrm{HOH}}}
\newcommand{\Gre}{\widetilde \Gamma'(\omega)}
\newcommand{\Gim}{\widetilde \Gamma''(\omega)}

\graphicspath{{./figs/}}

\begin{document}

\title{Time-dependent friction effects on vibrational infrared frequencies and line shapes of liquid water}

\author{Florian N. Br\"unig}
\affiliation{Freie Universit\"at Berlin, Germany}

\author{Otto Geburtig}
\affiliation{Freie Universit\"at Berlin, Germany}

\author{Alexander von Canal}
\affiliation{Freie Universit\"at Berlin, Germany}

\author{Julian Kappler}
\affiliation{Freie Universit\"at Berlin, Germany}

\author{Roland R. Netz}
\email{rnetz@physik.fu-berlin.de}
\affiliation{Freie Universit\"at Berlin, Germany}

\begin{abstract}
From ab initio simulations of liquid water, the time-dependent friction functions and time-averaged non-linear effective bond potentials for the OH stretch and HOH bend vibrations are extracted. The obtained friction exhibits adiabatic contributions at and below the vibrational time scales, but also much slower non-adiabatic contributions, reflecting homogeneous and inhomogeneous line broadening mechanisms, respectively. Compared to the gas phase, hydration softens both stretch and bend potentials, which by itself would lead to a red-shift of the corresponding vibrational bands. In contrast, non-adiabatic friction contributions cause a spectral blue shift. For the stretch mode, the potential effect dominates and thus a significant red shift when going from gas to the liquid phase results. For the bend mode, potential and non-adiabatic friction effects are of comparable magnitude, so that a slight blue shift results, in agreement with well-known but puzzling experimental findings. The observed line broadening is shown to be roughly equally caused by adiabatic and non-adiabatic friction contributions for both, the stretch and bend modes in liquid water. Thus, the understanding of infrared vibrational frequencies and line shapes is considerably advanced by the quantitative analysis of the time-dependent friction that acts on vibrational modes in liquids.
\end{abstract}

\maketitle

\begin{acronym}[Bash]
 \acro{aiMD}{ab initio Molecular Dynamics}
 \acro{ATR}{attenuated total reflection}
 \acro{DFT}{density functional theory}
 \acro{FPT}{first-passage time}
 \acro{GLE}{generalized Langevin equation}
 \acro{FTIR}{Fourier-transform infrared spectroscopy}
 \acro{FWHM}{full width at half maximum}
 \acro{IR}{infrared}
 \acro{LE}{Langevin equation}
 \acro{MD}{molecular dynamics}
 \acro{NM}{normal-mode}
 \acro{PMF}{potential of mean force}
 \acro{RC}{reaction coordinate}
\end{acronym}

\section{Introduction}

The OH stretch band in liquid water  is significantly red-shifted
and broadened compared
to the gas phase spectrum, while the HOH bend  frequency
is in fact slightly blue-shifted when going from gas to the liquid phase \cite{Falk1984}.
The broadening of the OH stretch band in liquid water is typically rationalized by a combination of homogeneous
and inhomogeneous effects  \cite{Bakker2010, Perakis2016}. Inhomogeneous line broadening is associated with
different hydrogen-bonding environments of individual OH bonds, which in the limit when the hydrogen-bonding pattern
changes more slowly than the OH vibrational period and in the presence of 
non-linearities in the OH bond potential,
 produce  vibrational frequencies  that vary  over time  \cite{Oxtoby1978,Moller2004,Auer2008}.
 Homogeneous line broadening reflects
the fast  coupling  of  OH bonds to their  neighboring water molecules, mostly via hydrogen bonding, which
reduces the vibrational life time since the vibrational energy is quickly
transported to neighboring molecules and thus  dissipated into collective 
modes.
Indeed, the vibrational life time of the OH stretch is very short (of the 
order of \SI{190}{fs}  \cite{Lock2002,Cowan2005})
and thus only  19  times longer than the OH-stretch vibrational period itself (of the order of \SI{10}{fs}).
The experimentally observed  red shift of
the OH stretch  band is usually rationalized by strong hydrogen bonding in liquid water, which extends
and thereby softens  the OH bond \cite{Bakker2010, Perakis2016}
(in fact, the relationship between the hydrogen-bond strength, the OH bond length  and the red shift of the stretch band
has been amply and partly controversially discussed in literature
\cite{Badger1934,Mikenda1986,Moller2004,Boyer2019}).
According to such reasoning,
the rather  small  frequency shift of the water bending mode when going from gas to liquid water
 could be argued to   imply  that the   bond angle potential  is only weakly perturbed
by the liquid water environment and thus that the coupling of  bend vibrations to the  hydration environment is weak.
This interpretation is puzzling though, since
the vibrational life time of the water bending mode in liquid water is  rather short
(around \SI{170}{fs} \cite{Ashihara2006,vanderPost2015})
 and thus  only
8.5 times longer than the vibrational period of \SI{20}{fs}, an even smaller ratio than for the stretch mode.
The short bend vibrational life time reflects
quick energy dissipation into librational modes \cite{Ashihara2007,Rey2009,Yu2020},
which in turn can be rationalized by efficient  multiphonon energy relaxation based on the excitation
of librational  overtones in liquids \cite{Ma2004}.
In this paper, we address the puzzle  posed by the different line shifts of the water stretch and bend modes
by analyzing the vibrational water dynamics  in terms of
the time-averaged non-linear bond potentials (as a function of the
bond length for the OH stretch and the bond angle for the HOH bend) and the  time-dependent friction functions,
which are extracted from extensive \ac{aiMD} simulations for 256 H$_2$O molecules.
In particular, we show that the slight blue shift of the water bend mode
when going from  gas to the liquid phase
 is not caused by a stiffening of the bend potential  \cite{Falk1984},
 but rather by the time dependence of the friction acting on  bending vibrations.

Time- or, equivalently, frequency-dependent friction arises whenever the dynamics of a many-particle system is
described in a low-dimensional reaction-coordinate space
 \cite{Lange2006,Horenko2007,Darve2009,Lesnicki2016,Deichmann2018,Jung2018,Meyer2019}
and its relevance for \ac{IR} spectra
was clearly demonstrated in the past
  \cite{ Metiu1977, Whitnell1992, Tuckerman1993,Gnanakaran1996,Joutsuka2011,Gottwald2015}.
All friction contributions that decay faster or similarly as the vibrational period stem from adiabatic solvent degrees of freedom
and account for dissipation into intra- and intermolecular degrees of freedom (including vibrational overtones)
\cite{Lawrence2003a,Ramasesha2013,Kananenka2018, Matt2018}, these friction contributions
dominate  the vibrational  energy relaxation  and lead to homogeneous line broadening.
Friction contributions that decay much slower than the vibrational time scale  describe the slowly changing  non-adiabatic
hydration environment
and in conjunction with non-linear bond potentials induce   inhomogeneous 
line broadening,
as our results explicitly demonstrate.
Of course, there is no clear-cut separation  between adiabatic and non-adiabatic solvent
relaxation modes \cite{Schmidt2005, DeMarco2016, Carpenter2017,Ojha2018}, 
prompting for a time-scale bridging  framework to treat
the dynamic coupling of molecular vibration modes  and their  environment.
In fact, the frequency-dependent friction function, which appears in the \ac{GLE},
is the  appropriate framework  to account for all these effects, with the 
only drawback that nuclear
quantum effects can at the current level of the  formalism not be included.
Only recent improvements of the extraction  methods  \cite{Daldrop2018b}  
allow to obtain these  time-dependent friction functions
 from \ac{aiMD} simulations and with high enough accuracy: this we self-consistently
demonstrate by predicting spectra from the \ac{GLE}  that are virtually indistinguishable  from the spectra directly obtained
from \ac{aiMD} simulations.

We find that the liquid environment in fact  significantly softens the time-averaged bond potentials, and it does so quite
similarly for the stretch and bend modes. Neglecting the frequency dependence of the friction,
both stretch and bend bands would thus be expected to be  red-shifted by comparable amounts
when going from gas to the liquid phase,
in stark contrast to the experimental finding  \cite{Falk1984}.
It turns out that  non-harmonic bond-potential effects are rather unimportant for the band position and thus cannot explain this puzzling finding.
Likewise, frequency-independent  friction  shifts the bands insignificantly and
 only increases  the line width, in agreement with expectations \cite{Oxtoby1978}.
 In contrast, the frequency dependence of the  friction  is crucial  and leads,  in conjunction with  non-linearities in the bond potentials,
 not only  to inhomogeneous  line broadening but
  also gives rise to pronounced blue shifts
  for both stretch and  bend bands. The mechanism for this blue shift is very general \cite{Metiu1977}, as we  analytically demonstrate.
The compensation of the potential red shift and the friction blue shift is incomplete for the stretch band but almost perfect for the bend band,
so the stretch band exhibits a significant net red shift
from gas to liquid, while the bend band only shows  a slight   blue shift 
in both experiments and simulations.
The absence of a significant frequency shift of the bend mode
does by no means  imply that bend vibrations couple less to their  environment than stretch vibrations
(as has been demonstrated previously \cite{Ashihara2007,Rey2009,Yu2020}):
rather,  it is  the balance of the potential and friction contributions to the  line shift, which both are caused
by interactions with the liquid  environment,  that is different for the stretch and bend bands.
We conclude that the coupling of water stretch and bend vibrations to other
 intra- and intermolecular degrees of freedom, as quantified by the time-averaged bond potentials and friction functions,
   is of  similar strength,
which explains their  similar vibrational life times, although their frequency shifts are rather  different,
which we rationalize by a subtle  difference of the  compensatory potential and friction effects.
The spectral  blue shift  due to frequency-dependent friction is a very general mechanism;
it transpires that  the concept of frequency-dependent friction is important for advancing the understanding of vibrational spectroscopy.

\section{System, Spectra  and Model}

\begin{figure}[htb]
\centering
\begin{overpic}[width=0.48\textwidth]{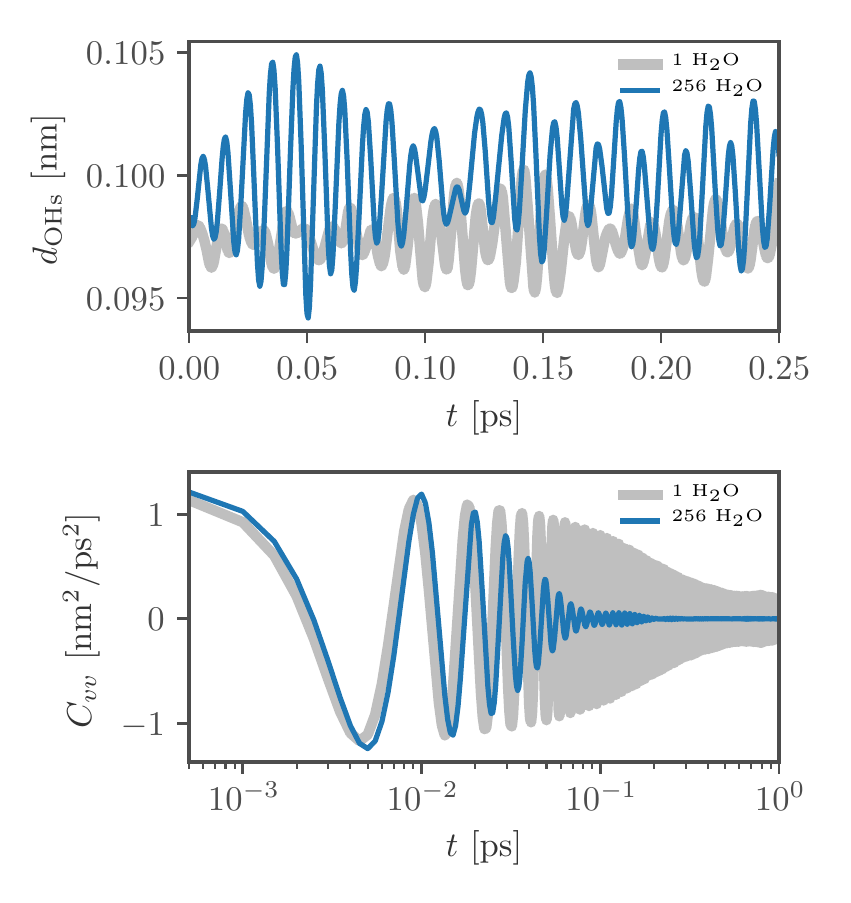}
\put(2,95){\large \bf A}
\put(2,45){\large \bf B}
\end{overpic}
\caption{{\bf A:} Trajectory of the  OH bond length, averaged over both OH bonds in a single water molecule,
from \acl{aiMD} (aiMD) simulations of one H$_2$O  in the gas phase
(grey line) and for
   256 H$_2$O  in the liquid phase  (blue line), both at \SI{300}{K}.
{\bf B:} Corresponding velocity autocorrelation functions.
}
\label{traject}
\end{figure}

We primarily  analyze \ac{aiMD} simulations of 256 H$_2$O (and D$_2$O for 
comparison)
molecules  in the liquid phase at \SI{300}{K} that neglect nuclear quantum effects.
Fig.~\ref{traject}A compares the trajectories of the mean  OH bond length 
 of a single
H$_2$O molecule
in  liquid H$_2$O (blue line) and in the
gas phase  (grey line), both at \SI{300}{K} (see Methods for simulation details).
The increase of the mean and the variance of the bond length in the liquid phase compared to the gas phase
 is clearly visible, which  reflects
the shift and softening of the OH bond potential due to
hydrogen bonding  in the liquid phase.
The slow fluctuations of the  oscillation amplitude
reflect   vibrational energy relaxations that occur over about 100 fs in the liquid phase (pure dephasing due to fluctuations of
the vibrational frequency  \cite{Stenger2001,Chuntonov2014} is  not easily visible in the time domain).
Similarly, the bond-length velocity autocorrelation function (VACF)  in fig.~\ref{traject}B demonstrates a significantly faster decay
and thus a decreased vibrational life time in the liquid phase.
Although the OH-stretch absorption spectrum is (apart from electronic and collective effects)
straightforwardly related to the OH bond-length  VACF  via Fourier transformation,
it turns out that only the careful analysis in terms of the \ac{GLE}    reveals the mechanisms
that determine  the OH-stretch vibration frequency and line shape.

\begin{figure*}[tb]
\centering
\begin{overpic}[width=\textwidth]{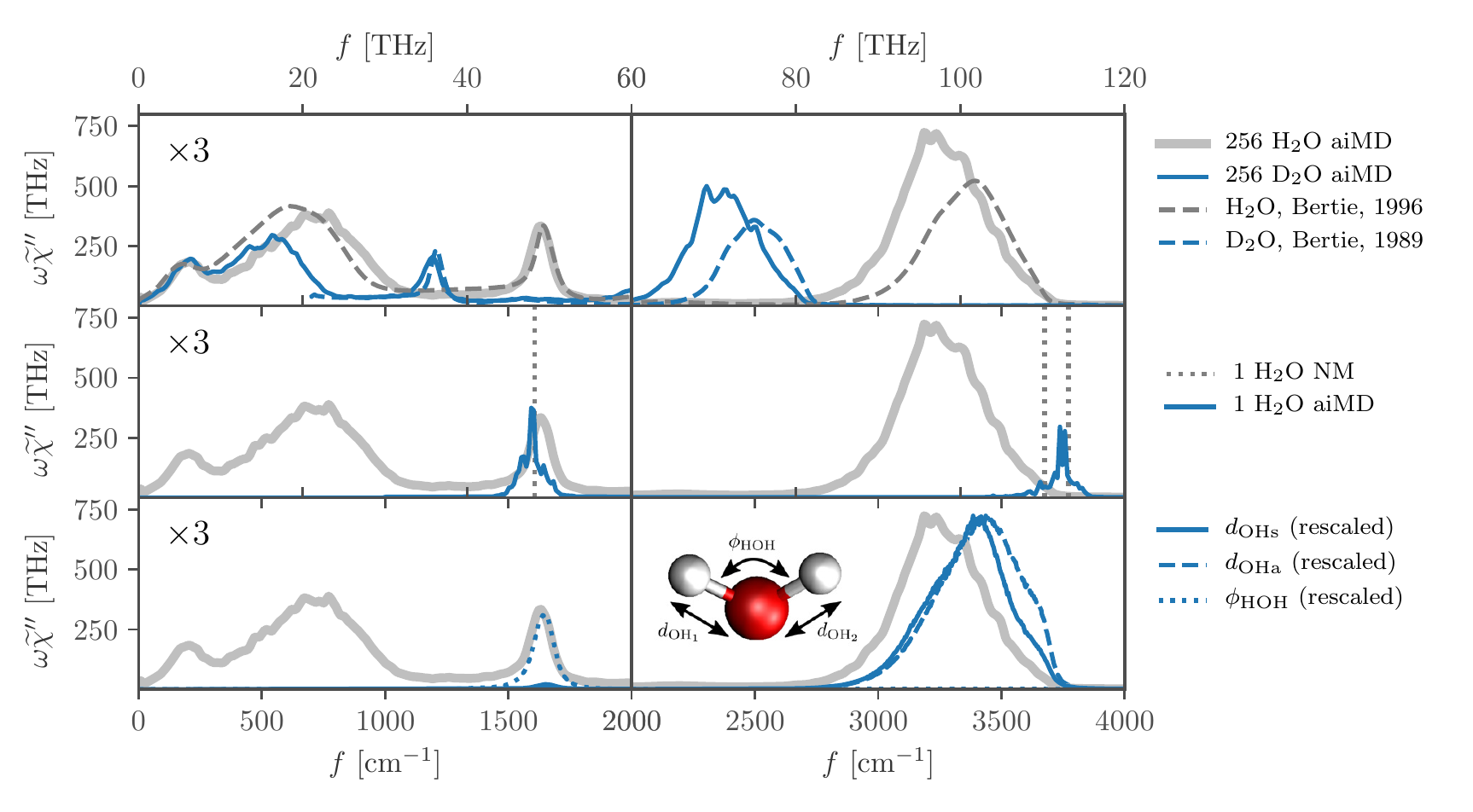}
\put(0,48){\large \bf A}
\put(0,33){\large \bf B}
\put(0,20){\large \bf C}
\end{overpic}
\caption{Absorption spectra of \ac{aiMD}  simulations at \SI{300}{K} of 256 H$_2$O are shown as grey solid lines in A-C.
The spectra up to \SI{2000}{cm^{-1}} are multiplied  by a factor three.
{\bf A:} Comparison to  \ac{aiMD} spectra   for liquid  D$_2$O  (blue solid line) and experimental data (obtained for \SI{298}{K}), shown as a grey broken line for H$_2$O \cite{Bertie1996} and a blue broken line for D$_2$O \cite{Bertie1989}.
{\bf B:} Comparison to  \ac{aiMD}  simulations of a single H$_2$O (blue solid line).
The  normal-mode frequencies of a single H$_2$O  are shown as vertical  
dotted lines.
{\bf C:} Comparison to  power spectra of the symmetric stretch, $\dohs$ (blue solid line),  anti-symmetric stretch, $\doha$ (blue broken line), and bend modes, $\hoh$ (blue dotted line),
 that are averaged over all molecules and rescaled to match the absorption spectrum.
 }
\label{oh_spec_intro}
\end{figure*}

Linear \ac{IR} spectroscopy experiments measure the absorbed power of light at  angular frequency $\omega=2\pi f$,
which is proportional to the imaginary part of the dielectric susceptibility $\widetilde \chi''(\omega)$.
Linear-response theory relates $\widetilde \chi''(\omega)$ to the total dipole-moment
autocorrelation (see SI section \ref{linearResponseSection}), allowing \ac{IR} spectra to be calculated  from equilibrium simulations
\cite{Silvestrelli1997,Heyden2010}.
Fig.~\ref{oh_spec_intro}A compares the \ac{IR} spectrum from \ac{aiMD} simulations of liquid  H$_2$O (grey solid line) and
 D$_2$O  (blue solid line) with  corresponding experimental data (grey and blue broken lines, respectively).
One discerns the stretch band (around \SI{3300}{cm^{-1}}  for H$_2$O and \SI{2400}{cm^{-1}} for D$_2$O in the \ac{aiMD} results)
and the bend band  (at \SI{1650}{cm^{-1}}  for H$_2$O and \SI{1200}{cm^{-1}}  for D$_2$O).
The librational band is produced by  a large number of different intermolecular  vibrational modes \cite{Schulz2018}
that are dominated  by rotational vibrations of water molecules
 in their hydrogen-bond environment (around \SI{700}{cm^{-1}} for H$_2$O and \SI{550}{cm^{-1}} for D$_2$O)
  and by  translational vibrations  of water molecules     against each other
 around \SI{200}{cm^{-1}} for both H$_2$O and D$_2$O.
 The  agreement  between \ac{aiMD} simulations, which fully account for electronic and nuclear polarizations, and experimental spectra is  good,
which suggests  that the chosen simulation method is well suited for modeling \ac{IR} spectra, although
the  agreement is known to be partly due to a cancellation of approximations in the employed density functional theory (DFT) and the
neglect of nuclear quantum effects \cite{Habershon2009,Marsalek2017}. However, the interplay of potential and frequency-dependent friction effects
we tackle in this paper  presumably is not  modified by nuclear quantum effects in a fundamental way, so that the conclusions we draw should
remain valid even beyond the Born-Oppenheimer approximation.

Fig.~\ref{oh_spec_intro}B compares  simulated  liquid H$_2$O   (grey) and 
single
H$_2$O  (blue solid line) spectra at \SI{300}{K}. The single water spectrum shows sharp peaks which perfectly coincide with
the normal mode frequencies  of a single water molecule (vertical dotted lines, computed on the same DFT level as the \ac{aiMD} simulations) at \SI{1607}{cm^{-1}}, \SI{3675}{cm^{-1}} and \SI{3772}{cm^{-1}},
which are  within \SI{20}{cm^{-1}} of the experimental values \SI{1594.7}{cm^{-1}}, \SI{3657.1}{cm^{-1}} and \SI{3755.9}{cm^{-1}} \cite{Fraley1969, McClatchey1973}.
Note that the OH-stretch band consists of  two modes, namely the low-frequency symmetric  mode, where both OH bonds
vibrate in phase, and the high-frequency anti-symmetric  mode, where the OH bonds vibrate out of  phase, which do not
clearly separate in the  liquid spectrum.
The  symmetric stretch mode in the gas phase
 shows a much smaller intensity than the anti-symmetric stretch mode,
 in agreement with  experiment \cite{McClatchey1973},
 which is caused by electronic polarization effects.
  The OH-stretch peak in the liquid is significantly red-shifted and enhanced  compared to gas phase,
  which is typically rationalized  by the softening of the OH bond potential and  the constructive
   collectivity of  OH-stretching vibrations in the liquid (see SI section \ref{collectivitySection}) \cite{Bakker2010, Perakis2016,Carlson2020};
   the significant enhancement
   is noteworthy, since one could expect the friction acting on the OH bond to be much stronger in the liquid and
   thus to reduce the vibrational amplitude.
   In contrast,  the HOH-bending mode  in the liquid is
slightly blue-shifted and not  enhanced,
   which can be  rationalized by  collective effects  that are slightly destructive (see SI section \ref{collectivitySection}).
 All these effects are fully accounted for   by the frequency-dependent friction acting on the different vibrational modes,
 as explained below.

The  vibrational  modes  of a water molecule can be described by the  bond
angle $\hoh$ and the
symmetric and anti-symmetric  stretch distances,   $\dohs = ({\doh}_1 + 
{\doh}_2)/2$ and $\doha = ({\doh}_1 - {\doh}_2)/2$,
where the two OH bond distances in a water molecule are denoted as ${\doh}_1$ and ${\doh}_2$,  all based on the nuclear positions
in the \ac{aiMD} simulations, as illustrated in the inset in fig.~\ref{oh_spec_intro}C.
 The  spectra of these three modes, averaged over all  water molecules in 
the liquid, are shown in fig.~\ref{oh_spec_intro}C
 ($\hoh$ as dotted, $\dohs$ as solid and $\doha$ as broken blue lines) and compared to the spectrum  from the
 total dipole moment.
 The agreement of the line frequencies and   shapes is quite good, except 
that
  the $\dohs$ and $\doha$ spectra are blue-shifted compared to the polarization spectrum. This blue shift is
  due to the neglect of spectral correlations between neighboring water molecules as discussed above and the neglect of electronic
  degrees of freedom in the nuclear-position mode spectra,
 as shown in the SI section \ref{atomCoordinatesSection}.
 The $\dohs$ and $\doha$ spectra overlap significantly, with a small red shift of  the $\dohs$ spectrum
relative to the $\doha$ spectrum, in accordance with previous observations \cite{Zhang2015}.
The  spectrum of the $\hoh$ mode overlaps perfectly with the   spectrum from the total (nuclear and electronic) dipole moment,
which is due to the fact that the bending angle vibrations of neighboring 
water molecules are only weakly (and in fact anti-) correlated,
as shown in the SI section \ref{collectivitySection}  \cite{Carlson2020}.
We conclude that the  spectra calculated from the total system polarization  (including nuclear and electronic polarization
from all water molecules and their correlations) match the spectra
based on the single-water nuclear-coordinate-based vibrational modes
rather faithfully, which is at the heart of the common molecular interpretation of IR spectra and also validates our further approach.

In the following, we will analyze the dynamics  of the water vibrational modes
based on  the one-dimensional  \ac{GLE}
\begin{align}
\label{eqM:gleah}
m \ddot{x}(t) = - \int_0^{t} \Gamma(t-t') \dot x(t') dt' - \nabla U[x(t)] + F_R(t)
\end{align}
which contains a in general  non-harmonic time-independent potential $U(x)$
that corresponds to a free energy as it results from integrating out all other degrees of freedom except $x(t)$.
The memory kernel $\Gamma(t)$
 describes the time-dependent friction acting on the fluctuating variable 
$x(t)$, which
 can be either the bond
angle $\hoh$, the
symmetric or the  anti-symmetric  stretch distances,   $\dohs$ or $\doha$.
The random force $F_R(t)$ has zero mean $\langle F_R(t) \rangle = 0 $ and fulfills the fluctuation-dissipation relation $\langle F_R(t) F_R(t') \rangle = k_BT \Gamma(t-t')$.
 Given a trajectory $x(t)$ from the \ac{aiMD} simulations, the effective mass $m$, the potential $U(x)$ and the friction function $\Gamma(t)$
 can be uniquely determined, as  described in SI sections \ref{FreeEnergyFitSection} and \ref{kernelExtractionSection}.
As a crucial test of the validity of the \ac{GLE}  and of our extraction methods, we will further below demonstrate that the \ac{GLE} accurately
reproduces the mode spectra calculated directly from the \ac{aiMD} simulations.

 For a harmonic potential, $U(x) = \frac{k}{2} x^2$,
the power spectrum can be given in closed form as
(see SI section \ref{gleResponseSection})
\begin{align}
\label{eqM:gleh_spec}
\omega \widetilde \chi''(\omega)  =  \frac{\omega^2\ \text{Re} \widetilde \Gamma(\omega)}{\left|k-m\omega^2 -
i \widetilde \Gamma(\omega)\omega\right|^2}
\end{align}
where the frequency-dependent friction is obtained by  a single-sided Fourier transform
$\widetilde \Gamma(\omega) = \int_{0}^{\infty} dt\ e^{ i \omega t} \Gamma(t)$.
In the limit of frequency-independent friction $\widetilde \Gamma(\omega) 
= \gamma$, this yields the
standard Lorentzian line shape \cite{Schrader1995} (see SI section \ref{hoResponseSection})
\begin{align}
\label{eqM:harm_spec}
\omega \widetilde \chi''(\omega)  =  \frac{\omega^2  \gamma } { (k-m \omega^2)^2  + \gamma^2 \omega^2 }
\end{align}
which  will be shown to give only a poor account of our simulated spectra.
Non-harmonic potentials  are parametrized as
\begin{align}
\label{eq:pot_param}
U(x) = \frac{k}{2} (x-x_0)^2 + \frac{k_3}{3} (x-x_0)^3 + \frac{k_4}{4} (x-x_0)^4
\end{align}
where $x_0$ is the position of the minimum of $U(x)$. Spectra in the presence of non-harmonic potentials
 are obtained from  numerical simulations of the \ac{GLE} using  a parametrized friction function of the form
 \cite{Marchesoni1982,Morrone2011,Lee2019}
\begin{align}
\label{eq:kernel_param}
\begin{split}
\Gamma(t) = &\sum_{i=0}^n \frac{\gamma_i}{\tau^e_i} e^{-t/\tau^e_i} \\
&+ \sum_{i=0}^l a_i e^{-t/\tau^o_i} \left[\cos(\omega_i t) + \frac{1}{\tau^o_i \omega_i} \sin(\omega_i t) \right]
\end{split}
\end{align}
consisting of $n$ exponentially decaying components with time scales $\tau^e_i$ and friction coefficients $\gamma_i$
as well as $l$ oscillating and decaying components with amplitudes $a_i$, 
oscillation frequencies $\omega_i$ and decay time scales $\tau^o_i$,
see SI sections \ref{KernelFitSection}--\ref{oscMemSection} for details.

\section{Results and Discussion}

\begin{figure*}[p]
\centering
\begin{overpic}[width=0.48\textwidth]{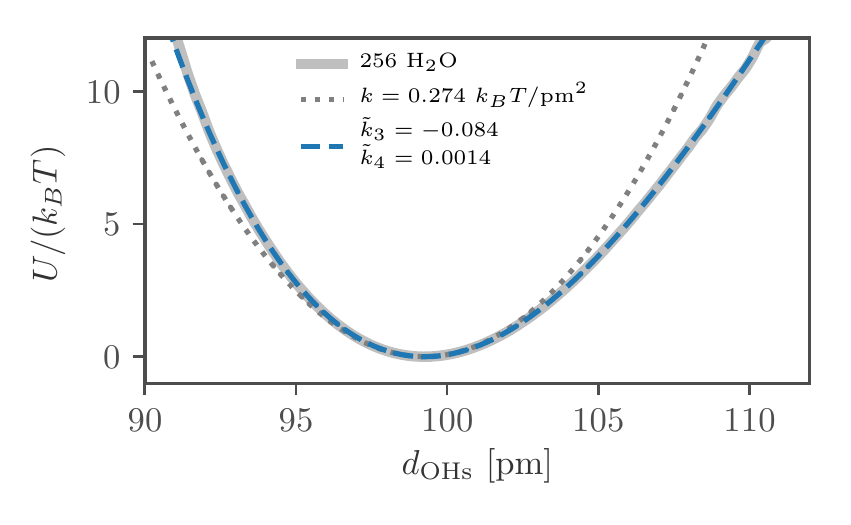}
\put(2,56){\large \bf A}
\end{overpic}
\begin{overpic}[width=0.48\textwidth]{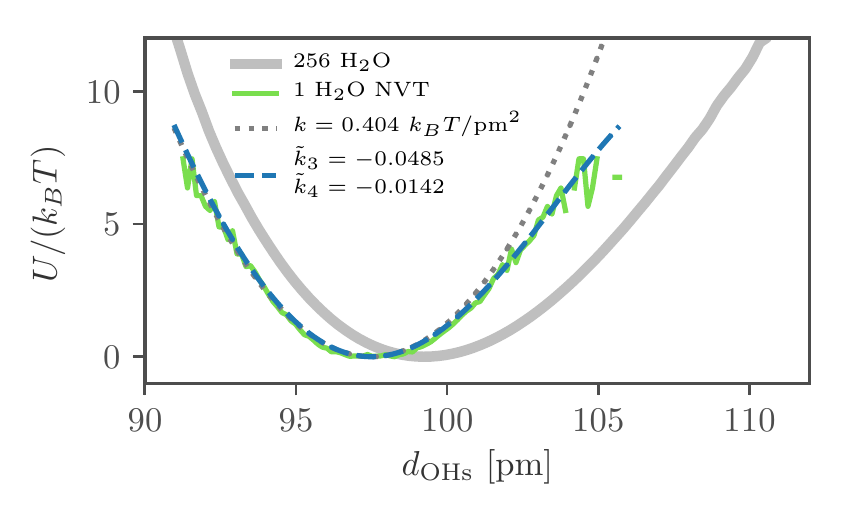}
\put(2,56){\large \bf B}
\end{overpic}
\begin{overpic}[width=0.48\textwidth]{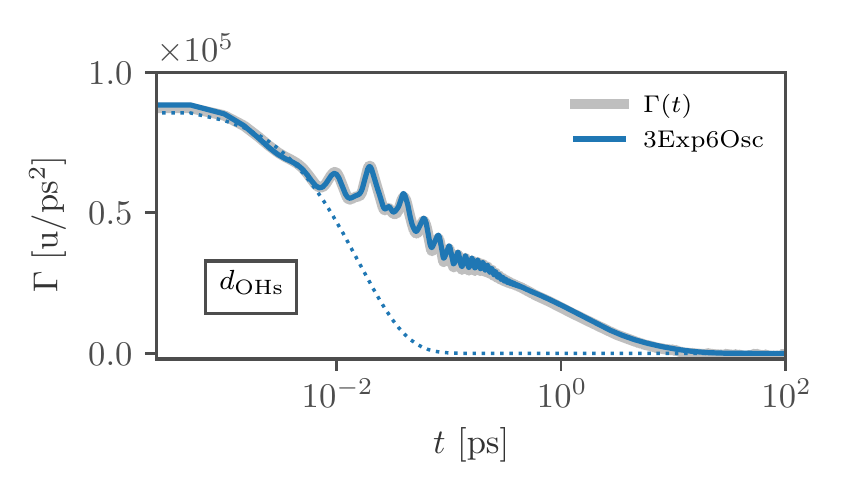}
\put(2,56){\large \bf C}
\end{overpic}
\begin{overpic}[width=0.48\textwidth]{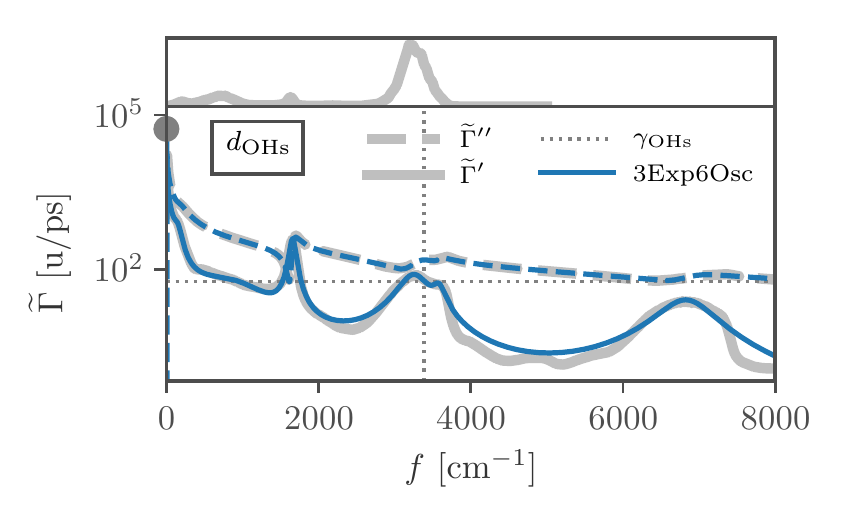}
\put(2,56){\large \bf D}
\end{overpic}
\begin{overpic}[width=0.48\textwidth]{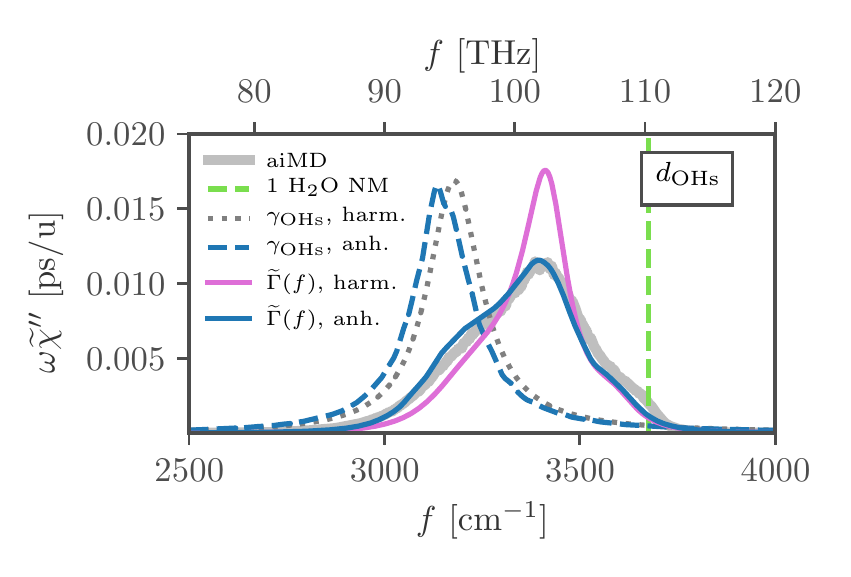}
\put(2,56){\large \bf E}
\end{overpic}
\begin{overpic}[width=0.48\textwidth]{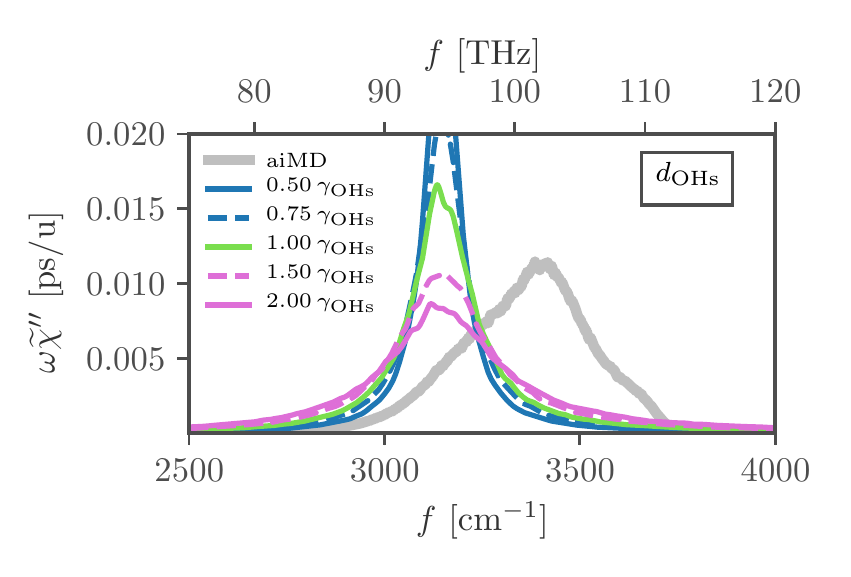}
\put(2,56){\large \bf F}
\end{overpic}
\caption{\footnotesize Results for the  symmetric stretch coordinate  $\dohs$  from \ac{aiMD} simulations.
{\bf A:} Potential $U(\dohs)$ for   256 H$_2$O in the liquid phase (grey solid line)
compared to the non-harmonic fit according to eq.~\eqref{eq:pot_param}  (blue broken line) and the harmonic part  (grey dotted line).
{\bf B:} Potential $U(\dohs)$ for  a single H$_2$O  (green solid line)  compared with the
non-harmonic fit according to eq.~\eqref{eq:pot_param}  (blue broken line) and the harmonic part  (grey dotted line), the liquid-phase
potential (grey solid line) is shown for comparison.
{\bf C, D:} Friction as a function of time and  frequency  (grey lines) compared with
the fit according to eq.~\eqref{eq:kernel_param} (blue lines). Real 
and imaginary parts in (D) are shown as solid and broken lines,
the spectrum on top is the full absorption spectrum from \ac{aiMD}.
The blue dotted line in (C) shows a single exponential with decay time $\tau=\SI{10}{fs}$, the dotted horizontal line
in (D) shows the constant  real friction $\gamma_{\rm OHs}= \widetilde{\Gamma}'(f_{\rm OHs})$ evaluated at the symmetric OH stretch vibrational frequency
$f_{\rm OHs}=\SI{3390}{cm^{-1}}$.
The grey circle denotes the static friction $\widetilde \Gamma'(0)$.
{\bf E:} Power spectrum $\omega \widetilde{\chi}''$
(grey solid line) compared to  models of varying complexity:
normal mode of single H$_2$O (broken vertical line),
Lorentzian with harmonic potential and constant friction $\gamma_{\rm OHs}$  (grey dotted line),
non-harmonic potential  and constant friction $\gamma_{\rm OHs}$  (blue broken line),
harmonic potential and frequency-dependent friction $\widetilde{\Gamma}(f)$ (purple solid line),
non-harmonic potential and frequency-dependent friction $\widetilde{\Gamma}(f)$  (blue solid line).
{\bf F:} Power spectrum $\omega \widetilde{\chi}''$    using the non-harmonic potential and different values of the constant friction $\gamma$, where $\gamma_{\rm OHs}= \widetilde{\Gamma}'(f_{\rm OHs} )$
 is the friction evaluated at the symmetric OH stretch vibrational frequency.
}
\label{ohs_friction}
\end{figure*}

\begin{figure*}[p]
\centering
\begin{overpic}[width=0.48\textwidth]{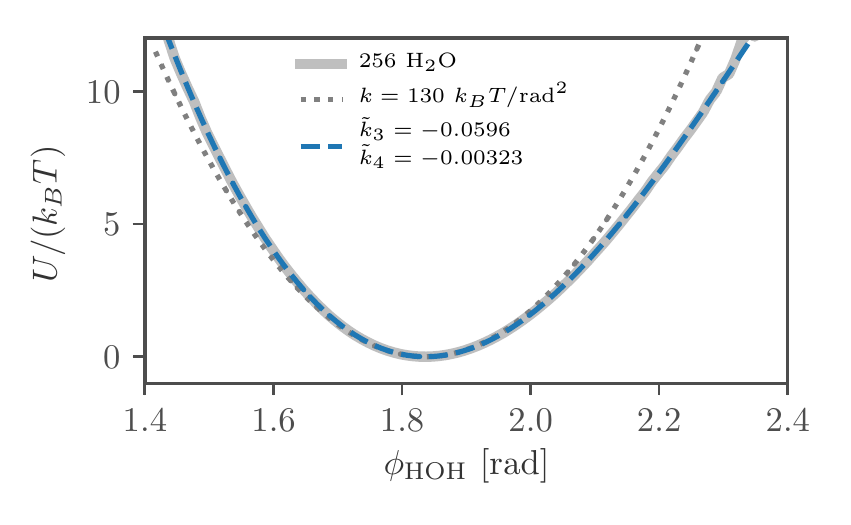}
\put(2,56){\large \bf A}
\end{overpic}
\begin{overpic}[width=0.48\textwidth]{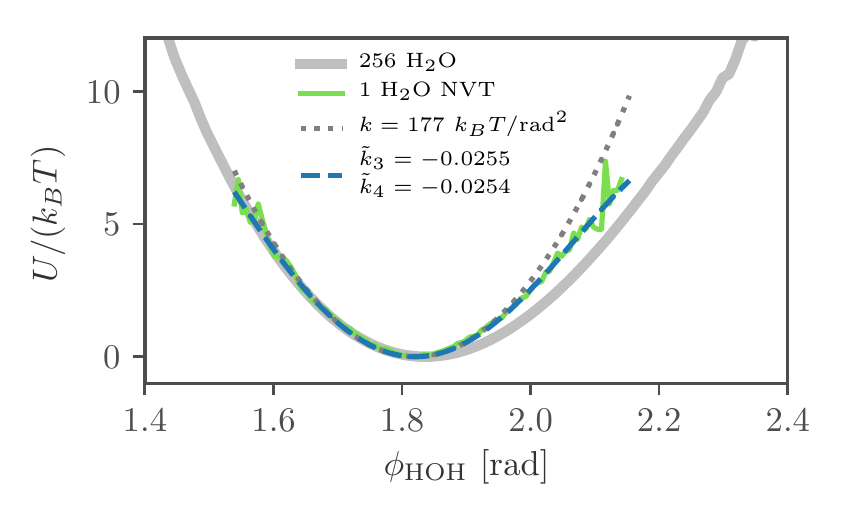}
\put(2,56){\large \bf B}
\end{overpic}
\begin{overpic}[width=0.48\textwidth]{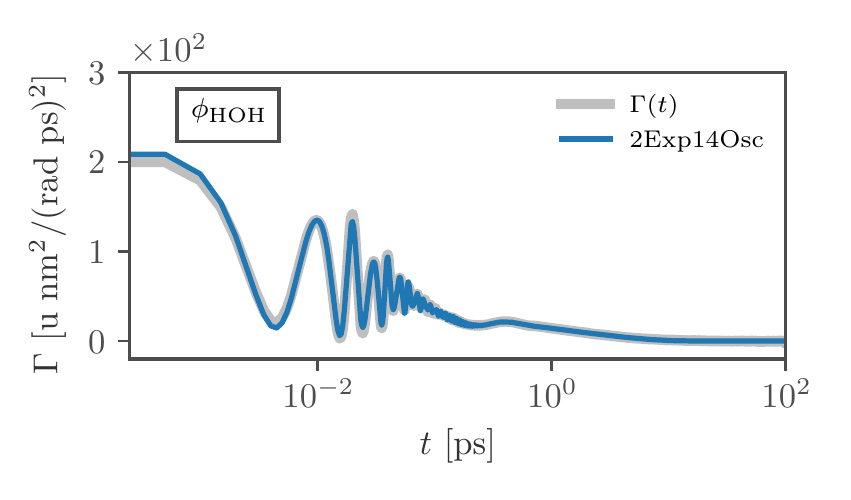}
\put(2,56){\large \bf C}
\end{overpic}
\begin{overpic}[width=0.48\textwidth]{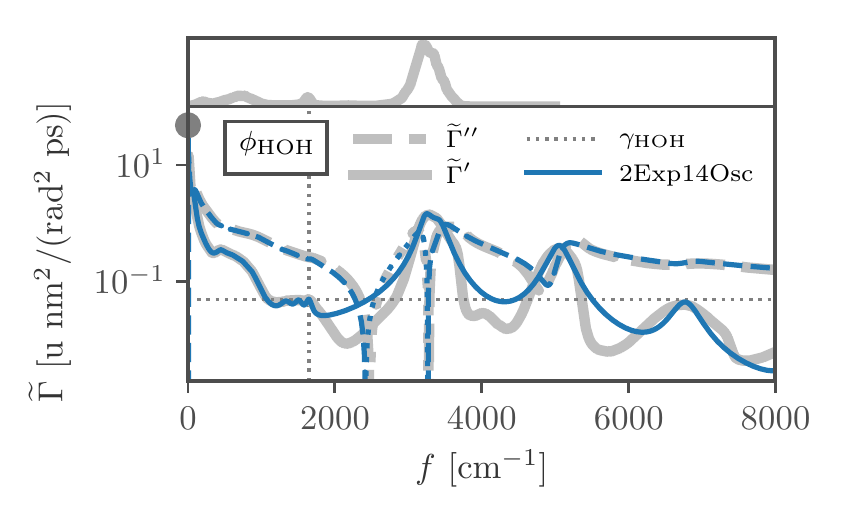}
\put(2,56){\large \bf D}
\end{overpic}
\begin{overpic}[width=0.48\textwidth]{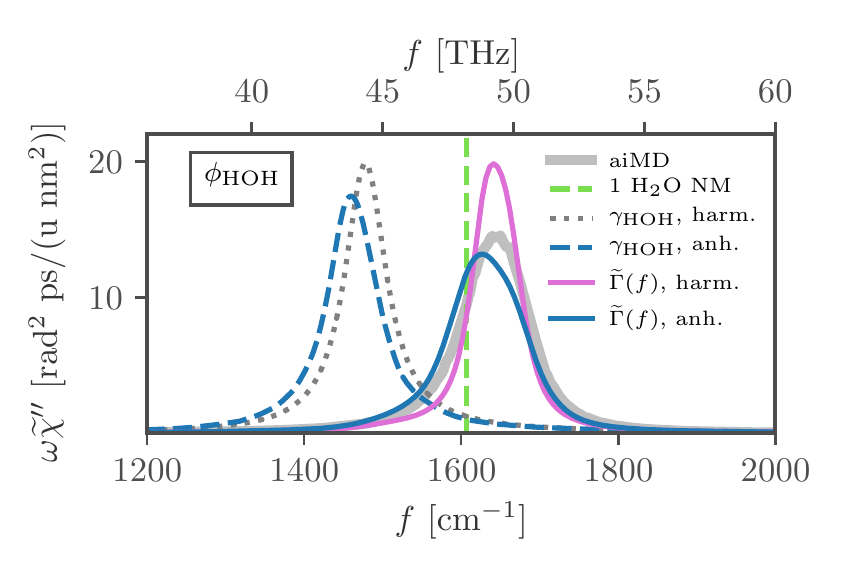}
\put(2,56){\large \bf E}
\end{overpic}
\begin{overpic}[width=0.48\textwidth]{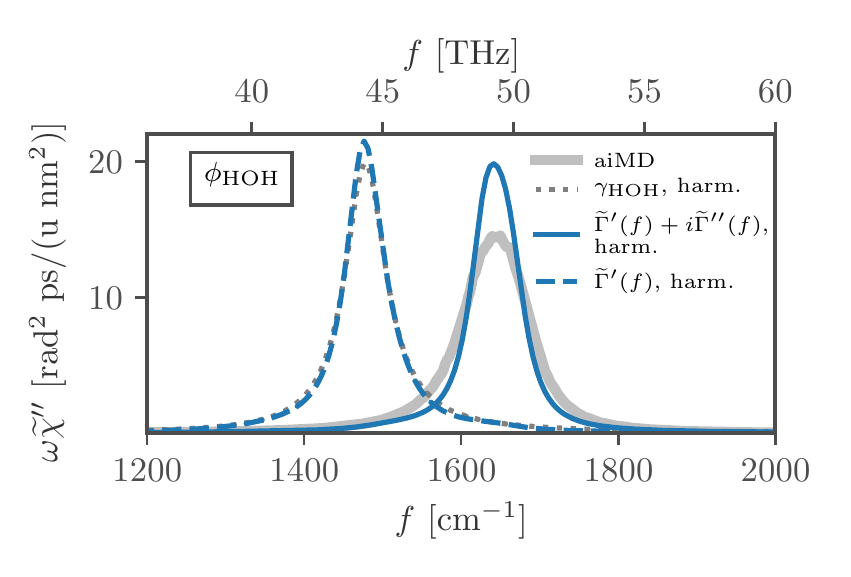}
\put(2,56){\large \bf F}
\end{overpic}
\caption{\footnotesize Results for the  bend coordinate  $\hoh$  from \ac{aiMD} simulations.
{\bf A:} Potential $U(\hoh)$ for 256 H$_2$O in the liquid phase (grey solid line)
compared to the non-harmonic fit according to eq.~\eqref{eq:pot_param}  (blue broken line) and the harmonic part  (grey dotted line).
{\bf B:} Potential $U(\hoh)$ for  a single H$_2$O  (green solid line)  compared with the
non-harmonic fit according to eq.~\eqref{eq:pot_param}  (blue broken line) and the harmonic part  (grey dotted line), the liquid-phase
potential (grey solid line) is shown for comparison.
{\bf C, D:} Friction as a function of time and frequency  (grey lines) compared with
the fit according to eq.~\eqref{eq:kernel_param} (blue lines). Real 
and imaginary parts in (D) are shown as solid and broken lines,
dotted lines denote negative values of the imaginary part, the spectrum on top is the full absorption spectrum from \ac{aiMD}.
The dotted horizontal line in (D) shows the constant  real friction
$\gamma_{\rm HOH}= \widetilde{\Gamma}'(f_{\rm HOH})$ evaluated at the bend  vibrational frequency
$f_{\rm HOH}=\SI{1650}{cm^{-1}}$. The grey circle denotes the static friction $\widetilde \Gamma'(0)$.
{\bf E:} Power spectrum $\omega \widetilde{\chi}''$
(grey solid line) compared to  different models:
normal mode of single H$_2$O (broken vertical line),
Lorentzian with harmonic potential and constant friction $\gamma_{\rm HOH}$  (grey dotted line),
non-harmonic potential  and constant friction $\gamma_{\rm HOH}$  (blue broken line),
harmonic potential and frequency-dependent friction $\widetilde{\Gamma}(f)$ (purple solid line),
non-harmonic potential  and frequency-dependent friction $\widetilde{\Gamma}(f)$  (blue solid line).
{\bf F:} Power spectrum $\omega \widetilde{\chi}''$    using the harmonic 
 potential part
 and the constant friction $\gamma_{\rm HOH}$  (grey broken line),
 the real frequency-dependent friction only $\widetilde{\Gamma}'(f)$ (blue broken line),
 the   real and imaginary  frequency-dependent friction  $\widetilde{\Gamma}'(f)+i \widetilde{\Gamma}''(f)$ (blue solid line).
 The grey solid line is the spectrum from \ac{aiMD} simulations.
 }
\label{hoh_friction}
\end{figure*}

We start with a discussion of the symmetric stretch mode $\dohs$.
The potential  (or rather the free energy) $U(\dohs)$
  from the \ac{aiMD} simulations for liquid water  (grey solid line) is  in fig.~\ref{ohs_friction}A
compared  with a non-harmonic fit according to eq.~\eqref{eq:pot_param} (blue broken line),
the harmonic contribution is shown as a  grey dotted line.
The  comparison of the  liquid and gas-phase bond  potentials in fig.~\ref{ohs_friction}B shows that the minimum of the potential
(i.e. the most probable OH bond length)  increases
 from $x_0=\SI{97.50}{pm}$ in the gas phase  to $x_0=\SI{99.25}{pm}$ in the liquid;
 at the same time  the harmonic force constant decreases from $k/k_BT=\SI{0.404}{pm^{-2}}$ in the gas phase
  to $k/k_BT=\SI{0.274}{pm^{-2}}$ in the liquid.
  This  softening of the potential is
due to elongation of the bond, caused by hydrogen bonding in the liquid, and will be shown to induce a pronounced spectral red shift.
 Furthermore, the potential  non-harmonicity increases, as can be seen by 
comparing the reduced cubic
  potential coefficient in the liquid phase  $\tilde k_3 = k_3/k_BT (k/k_BT)^{-3/2} = -0.0840$ with the
   value in the gas phase $\tilde k_3 = -0.0485$.

The time-dependent friction function  for the symmetric stretch mode
$\Gamma(t)$ extracted from  \ac{aiMD} simulations, grey line in fig.~\ref{ohs_friction}C,
shows multi-exponential decay characterized  by relaxation times from a few fs to many ps, which is appreciated by comparison with a single-exponential
function with decay time $\tau=\SI{10}{fs}$ (dotted blue line,  the logarithmic time axis should be noted).
This in particular means that  $\Gamma(t)$ accounts for solvent relaxations that are fast  (adiabatic)
and slow (non-adiabatic) with respect to the OH vibrational period of \SI{10}{fs};
one thus expects homogeneous as well as inhomogeneous line broadening to occur, as indeed borne out by our analysis below.
The oscillations that appear in $\Gamma(t)$ at around  \SIrange{10}{250}{fs}   reflect the dissipative coupling of  symmetric stretch vibrations
to anti-symmetric stretch as well as  higher-harmonic bend  and librational  modes. This is  illustrated  by
the real and imaginary frequency-dependent friction components
$\Gre + i \Gim = \int_{0}^{\infty} dt\ e^{i \omega t} \Gamma(t)$
 in fig.~\ref{ohs_friction}D (solid and broken grey lines, respectively),
 which exhibit maxima at the OH-stretching  and  HOH-bending frequencies and also  at  their higher harmonics.
 The friction function thus   details the vibrational energy dissipation
 of a given vibrational mode within
a water molecule as well as with the surrounding water
 and in particular  accounts for Fermi resonances between different vibrational modes
(albeit on the classical level).
Also non-Condon effects, which arise due to modifications of the transition dipole moment of a vibrational mode
due to time-dependent changes of the solvation environment of a molecule \cite{Schmidt2005}, are   included
via the interplay of the potential $U(\dohs)$ and the time-dependent friction function  $\Gamma(t)$.
Interestingly, the symmetric stretch shows a much stronger frictional damping at the characteristic frequency of
the  bending mode than the anti-symmetric stretch mode, shown in  SI section \ref{ohDecompSection}, which points to a stronger dissipative coupling of  bending vibrations
with symmetric than
with anti-symmetric stretch vibrations.
For simulations of the \ac{GLE}, which are necessary for the analysis of the coupling between  non-linearities in the potential and
frequency-dependent friction, we fit  $\Gre$ by the expression eq.~\eqref{eq:kernel_param}
with   a sum of three exponential  and
six oscillating functions, see SI section \ref{KernelFitSection} for details. The fit  shown in blue in figs.~\ref{ohs_friction}C and D describes the simulated friction function
equally well in the time as well as in the frequency domain.

The vibrational spectrum of the $\dohs$ mode  directly extracted from \ac{aiMD} simulations is shown in fig.~\ref{ohs_friction}E as a grey solid line.
The simplest possible model for a vibrational line shape is
the Lorentzian model eq.~\eqref{eqM:harm_spec} for a harmonic potential and a constant, frequency-independent friction.
Using  $k/k_BT=\SI{0.274}{pm^{-2}}$ from the harmonic  fit in fig.~\ref{ohs_friction}A  and the  friction
$\gamma_{\rm OHs}= \widetilde{\Gamma}'(f_{\rm OHs})$ in fig.~\ref{ohs_friction}D at the stretch vibrational frequency $f_{\rm OHs}=\SI{3390}{cm^{-1}}$, we obtain the grey dotted line in fig.~\ref{ohs_friction}E. Compared to the normal-mode frequency of the gas phase,
denoted by a vertical green broken line, the Lorentzian is significantly red-shifted by about \SI{500}{cm^{-1}}, the width of the
Lorentzian reflects homogeneous line broadening due to adiabatic solvent friction  that is described by the frequency-independent
constant $\gamma_{\rm OHs}$.
Note that the Lorentzian is  considerably red-shifted and narrower compared
to the spectrum extracted from the \ac{aiMD} simulation (grey line). Interestingly, the friction  $\gamma_{\rm OHs}$ that acts at the vibration frequency
$f_{\rm OHs}$ is about two orders of magnitude smaller than the friction in the static limit $f=0$, as seen in fig.~\ref{ohs_friction}D,
which explains why the stretch vibrational dynamics shown in fig.~\ref{traject}A is rather weakly damped.
The power spectrum in the presence of  the  full non-harmonic potential $U(\dohs)$ and constant friction
$\gamma_{\rm OHs}$,
obtained  from numerical simulations of the regular Langevin equation  (blue broken line,  see SI section \ref{GLESimulationSection} for details),
is  only slightly red-shifted with respect to the Lorentzian obtained for 
a harmonic potential, which is
expected based on perturbation theory \cite{Oxtoby1978}.
 We conclude that non-linearities in the potential have for constant friction  only an insignificant influence on the line frequency and shape.
 The peak frequency  of a  Lorentzian does not depend on the value of the 
constant friction $\gamma$
 (see SI section \ref{deltaExampleSection}),
 which is approximately true
  also in the presence of the non-harmonic potential
 $U(\dohs)$, as demonstrated in fig.~\ref{ohs_friction}F where spectra from
  numerical simulations for varying $\gamma$ are compared.
  We next check for the influence of  time-dependent friction on the spectrum.
For a harmonic potential and for frequency-dependent friction,
the spectrum is determined analytically by eq.~\eqref{eqM:gleh_spec} and shown in fig.~\ref{ohs_friction}E as a purple
 solid line. A significant blue shift compared to the results for constant friction is obtained,
  so that the position of the spectrum agrees very well with the simulated spectrum, while the line shape
 is too narrow. The blue shift can be understood based on simple and rather general analytic arguments, as shown below.
The spectrum obtained from the \ac{GLE} in the presence of the non-harmonic potential  $U(\dohs)$ and time-dependent friction $\Gamma(t)$,
shown by the blue solid line in fig.~\ref{ohs_friction}E
(here numerical simulations are employed), is significantly broadened compared to the results obtained
for a harmonic potential and time-dependent friction $\Gamma(t)$ (purple line).
This  reflects the effects of inhomogeneous line broadening  \cite{Oxtoby1978},
 and reproduces the spectrum
 extracted from the \ac{aiMD} simulations (grey line) almost  perfectly; in fact,  inhomogeneous line broadening is quite substantial and accounts 
for \SI{52}{\%} of the total line broadening.
 This means that the \ac{GLE}, when used in conjunction with the properly extracted
 non-harmonic time-averaged
 potential $U(\dohs)$ and time-dependent friction $\Gamma(t)$, reproduces 
the system dynamics very well, which is not guaranteed in general
 since the projection onto the \ac{GLE} neglects non-linear friction effects \cite{Gottwald2015}.

The $\hoh$ water bending coordinate is analyzed analogously:
The bend angle potential $U(\hoh)$ in  fig.~\ref{hoh_friction}A extracted 
from  \ac{aiMD} simulations (grey line)
includes  significant non-linear contributions as appreciated by a comparison of the non-harmonic fit (blue broken line) with the
harmonic part (dotted line) and as witnessed by the magnitude of the  reduced cubic and quartic fit parameters $\tilde k_3 = -0.0596$ and $\tilde k_4 = k_4/k_BT (k/k_BT)^{-2}= -0.00323$. 
Different from the situation for the stretch potential, the liquid environment shifts the most probable bending angle only very slightly. The potential is softened considerably,
as is seen by a comparison of the shape and fit parameters of the gas and liquid phase potentials $U(\hoh)$  in fig.~\ref{hoh_friction}B,
which can be rationalized by the fact that attractive electrostatic interactions, which are predominant for strongly correlated polar liquids such as water,
exhibit negative curvature throughout their entire interaction range.
The   time-dependent friction $\Gamma(t)$  extracted from the simulations 
in fig.~\ref{hoh_friction}C
(grey line) shows a  broad  decay but  more pronounced
oscillations compared to  the stretch vibrations in fig.~\ref{ohs_friction}C.
The fit (blue solid  line) to the simulated
real frequency-dependent friction $\widetilde{\Gamma}'(f)$ (grey solid line)
in fig.~\ref{hoh_friction}D requires two exponential and 14 oscillatory functions to describe the simulated data satisfactorily, see SI section \ref{KernelFitSection} for details.
The dissipative damping is  significantly more pronounced at stretch frequencies around  \SI{3400}{cm^{-1}}
and at the overtones of the bending around  \SI{3300}{cm^{-1}} and  around  \SI{4950}{cm^{-1}} than at the bending fundamental  around  \SI{1650}{cm^{-1}} itself, indicative of the non-linear coupling between different modes
 and overtones (where it should be noted that coupling of bend vibrations 
to higher-frequency modes and overtones
 are reduced when quantum effects are properly included  \cite{Rey2009}).

The vibrational  spectrum of the $\hoh$ coordinate  from the \ac{aiMD} simulations is shown in fig.~\ref{hoh_friction}E as a grey solid line
and is   weakly blue-shifted from the gas phase normal mode (vertical green broken line), which is a surprising fact and will
be explained now by compensatory potential and friction effects.
The  spectrum from the Lorentzian model eq.~\eqref{eqM:harm_spec} (grey dotted line) using only  the  harmonic potential part  of $U(\hoh)$
and the  frequency-independent friction
$\gamma_{\rm HOH}=\widetilde{\Gamma}'(f_{\rm HOH})$, obtained  at the bending peak at $f_{\rm HOH}=\SI{1650}{cm^{-1}}$
(horizontal broken line in  fig.~\ref{hoh_friction}D),
 is significantly red-shifted and is not modified much by including the non-harmonic potential contributions
(blue broken line).
Including the complex frequency-dependent friction $\widetilde{\Gamma}(f)$  but only the harmonic part of $U(\hoh)$
the purple line is obtained, which is blue  shifted with respect to the constant-friction case and reaches
the frequency of the simulated curve but is too narrow.  Including
the complex frequency-dependent friction $\widetilde{\Gamma}(f)$
and also  the full non-harmonic  potential $U(\hoh)$,
 the \ac{GLE}  (indicated by the blue line) rather accurately reproduces the position and width of the simulated spectrum.
In agreement with our stretch-vibration results in fig.~\ref{ohs_friction}E,
we detect considerable  inhomogeneous line broadening (amounting to \SI{47}{\%} of the total line broadening) from the comparison of the results
with and without non-harmonic potential contributions in the presence of  
frequency-dependent friction.
In contrast to the
stretch-vibration results, we see that the blue shift induced by including the frequency dependence
of the friction almost exactly cancels the red shift due to the softening 
of the bond potential in the liquid phase, which means that the frequency dependence of $\widetilde{\Gamma}(f)$ close to the characteristic bend-mode frequency is more pronounced compared to the stretch mode.

It turns out that the imaginary and real parts of the frequency-dependent 
friction
influence  the line position and shape quite differently \cite{Metiu1977}.
This  is illustrated in fig.~\ref{hoh_friction}F by comparing spectra using only the harmonic part of the potential
for constant friction (grey dotted line),  for purely real frequency-dependent friction
 $\widetilde{\Gamma}'(f)$
 (blue broken line)
 and for friction that contains both real and imaginary
frequency-dependent parts $ \widetilde{\Gamma}'(f) + i \widetilde{\Gamma}''(f)$ (blue solid line), note that for purely imaginary friction the spectrum according to eq.~\eqref{eqM:gleh_spec} exhibits a singularity and thus is not shown.
It is in fact the imaginary part $\widetilde{\Gamma}''(f)$
that gives rise to the blue shift, as is now  explained by a simple analytical argument.

For this we consider a single-exponential memory function $\Gamma(t)=\gamma \tau^{-1} \exp(-t/\tau)$.
The single-sided Fourier transform  is given as
$\widetilde \Gamma(\omega) = \int_{0}^{\infty} dt\ e^{i \omega t} \Gamma(t)= \gamma/(1 - i \tau \omega)$
with the asymptotic limits $\widetilde \Gamma(\omega)  \simeq \gamma(1 + i \omega \tau)$ for small $\omega$ and
$\widetilde \Gamma(\omega) \simeq  i \gamma/(  \omega \tau)$ for large $\omega$, both deviations
from the zero-frequency limit $\widetilde \Gamma(\omega \to 0)  \simeq \gamma$ turn out to be imaginary,
 which already hints at why the imaginary part of the friction determines 
the line position, as demonstrated  in fig.~\ref{hoh_friction}F.
A general form that contains both asymptotic limits is given by
$\widetilde \Gamma(\omega) \simeq \gamma + i a \omega +  i b/\omega$, where
$a= \gamma \tau$ and $b=0$ for small $\omega$ and $a=0$ and $b=\gamma/\tau$ for large  $\omega$.
By inserting this asymptotic form  into eq.~\eqref{eqM:gleh_spec},
the  Lorentzian line shape eq.~\eqref{eqM:harm_spec} is recovered but with an effective
mass $m_{\rm eff}= m-a$ and  an effective potential curvature $k_{\rm eff}=k+b$.
The vibrational  frequency turns out to be
\begin{equation}
\omega_0=\sqrt{\frac{k_{\rm eff}}{m_{\rm eff}}}=  \sqrt{\frac{k+b }{m-a}}
 \end{equation}
 and in fact increases  both in the small and large frequency limits, since $a$ and $b$ are positive constants for single-exponential
 memory.
 Thus, a  blue shift  of the vibrational frequency
  is very generally expected for frequencies  where the frequency-dependent friction is described by the asymptotic  form
  $\widetilde \Gamma(\omega) \simeq \gamma + i a \omega +  i b/\omega$ with positive $a$ and $b$. In fact, this functional form  is  able  to describe  the stretch and band friction functions rather accurately
   around the stretch and band frequencies, respectively,
  as inspection of figs.~\ref{ohs_friction}D  and ~\ref{hoh_friction}D shows.

 The full width at half maximum of a Lorentzian is given  as
 $\gamma/m_{\rm eff}\simeq\gamma/(m-a)$;  thus  the line width is, within 
the harmonic approximation,  predicted to
 slightly increase for the stretch  band (since $\widetilde \Gamma''(f)$  
slightly increases at the stretch vibrational frequency
 in  fig.~\ref{ohs_friction}D  and thus
 $a$ is positive) but to stay rather constant for the bend band
 (since  $\widetilde \Gamma''(f)$ slightly decreases   at the stretch vibrational frequency   fig.~\ref{hoh_friction}D
  and thus  $a$ presumably is small
 and dominated by $b$). These predictions  are in good agreement with the 
results shown in
  figs.~\ref{ohs_friction}E  and \ref{hoh_friction}E  for the scenario  of a harmonic potential and friction-dependent friction
  (purple lines).
 Clearly, the exact line shape and position are  determined by the interplay of  non-linearities  of the potential
 and frequency-dependent friction,
 but the simple harmonic model discussed here allows to appreciate part of the mechanisms at play.

\section{Conclusions}

While  frequency-independent friction, which  reflects the  fast adiabatic dissipative channels available
for a specific vibration,  mostly modifies  the line width via homogeneous line broadening, but not the line position,
which holds approximately even in the presence of non-linear potential contributions,
the full frequency dependence of the friction,
which in particular  accounts for the slower solvent relaxation processes,
gives rise to a blue shift and additional line broadening,  the latter reflects what is typically called inhomogeneous line broadening.
In contrast,
softening of the bond potential in the liquid environment, which is due to hydrogen bonding and hydration interactions,
gives rise to a red shift.
So we find  that the line shapes and positions of the bend and stretch bands in liquid water can be understood by
a careful discussion of the compensatory effects of
frequency-dependent friction and harmonic as well as  non-harmonic potential contributions.
For  stretch vibrations, the  bond softening dominates and therefore the stretch vibration is red-shifted when going from
gas to liquid water; for bend vibrations the potential-induced red shift and the friction-induced blue shift almost exactly compensate.
This of course does not imply that the coupling of bend vibrations to the hydrating liquid environment  is weaker than for stretch vibrations, as one might naively guess from only looking at the frequency  shifts, rather the contrary is true.
 It turns out that it is the imaginary part of the frequency-dependent friction that gives rise to the blue shift,
 in line with previous arguments \cite{Metiu1977}.
The situation is rather complex, though, since the effects due to the  frequency-dependency of the friction and  due to non-linearities in the potential do not decouple. 
Our methodology is different from previous approaches to describe the infrared line shapes of water \cite{Auer2008, Ni2015}, since we deploy the time-averaged bond potential as it is defined in the \ac{GLE}. This in particular means that in our approach, inhomogeneous line broadening enters via the time-dependent friction function, not via a time-dependent bond potential.

As mentioned before,
our ab initio simulations  neglect nuclear  quantum effects \cite{Habershon2009,Marsalek2017},
owing to the fact that methods to extract friction functions from
path-integral simulations are not yet available. This approximation presumably is permissible in the present context,
as we target the general  compensatory effects the liquid environment has 
on
bond potentials and the bond friction function, which should not be fundamentally changed by nuclear quantum effects.
In the future, it would be interesting to extract \ac{GLE} parameters  from path integral simulations \cite{Habershon2009,Marsalek2017} and from mixed quantum/classical approaches \cite{Lawrence2003a,Medders2015,Kananenka2018}.

\section{Methods}

All Born-Oppenheimer \ac{aiMD} simulations were performed with the CP2K 4.1 software package  using a double-$\zeta$ basis set for the valence electrons, optimized for small molecules and short ranges, (DZVP-MOLOPT-SR-GTH), dual-space pseudopotentials, the BLYP exchange-correlation functional,
D3 dispersion correction
and a  cutoff for the plane-wave representation set to \SI{400}{Ry} \cite{Hutter2014, VandeVondele2007,Grimme2010}. The \ac{aiMD} simulations were performed using a time step of \SI{0.5}{fs} under NVT conditions at $\SI{300}{K}$ by coupling all atoms to a CSVR thermostat with a time constant of \SI{100}{fs} \cite{Bussi2007}. The bulk systems contain 256 molecules subject to periodic boundary conditions in a cubic cell of size $(\SI{1.9734}{nm})^3$, corresponding to densities of \SI{996.4}{kg/m^3} for H$_2$O and 
\SI{1107.8}{kg/m^3} for D$_2$O.
The total trajectory lengths of the liquid systems are \SI{230}{ps} for H$_2$O and \SI{130}{ps} for D$_2$O. Simulations of single H$_2$O, representing the gas phase data, were performed in the NVE ensemble with 47 initial configurations sampled from a \SI{25}{ps} NVT simulation using an individual thermostat with a time constant of \SI{10}{fs} for each atom. The NVE simulations were each run for \SI{10}{ps} with a time step of \SI{0.25}{fs}. The distributions of their initial configurations sample well the 
equilibrium distributions as shown in SI section \ref{initDistSection}.

Linear response theory relates the dielectric susceptibility $\chi(t)$ to 
the equilibrium autocorrelation of the dipole moment $C(t)=\langle \bm{p} (t)\bm{p}(0)\rangle$, reading in Fourier space \cite{Kubo1957}
\begin{align}
\label{linearResponseFT}
\widetilde \chi(\omega) = \frac{1}{V \epsilon_0 k_BT}\left( C(0) - i \frac{\omega}{2} \widetilde C^+(\omega) \right)
\end{align}
with system volume $V$, thermal energy $k_BT$ and vacuum permittivity $\epsilon_0$. \ac{IR} spectra can therefore be calculated straight-forwardly 
from sufficiently long trajectories from \ac{aiMD} simulation data using eq.~\eqref{linearResponseFT} and the Wiener-Kintchine relation \cite{Wiener1930, Carlson2020}, derived in SI section \ref{WienerKintchineSection}. 
Quantum corrections have previously been addressed, but were not applied here \cite{Ramirez2004}. The molecular dipole moments are obtained after Wannier-center localization of the electron density at a time resolution of \SI{2}{fs}. The Wannier centers were assigned to the molecule of the nearest oxygen, which always results in exactly four Wannier centers per water molecule. A charge of \SI{-2}{e} is assigned to each Wannier center, 
which together with the nuclear charges, reduced by the electronic charges of the inner shells, allows for the calculation of the dipole moment. The power spectra are smoothed using a Gaussian kernel with width \SI{10}{cm^{-1}}.
The normal mode analysis was performed for an energetically minimal configuration of a single H$_2$O using the implementation in CP2K 4.1 and the same ab initio model as for the \ac{aiMD} simulation.

\begin{acknowledgments}
We gratefully acknowledge support by the DFG grants SFB 1078 and SFB 1114, the MaxWater 
initiative from the Max Planck Society,
the ERC Advanced Grant grant agreement No. [835117],
computing time on the HPC cluster at ZEDAT, FU Berlin and the computational resources provided by the North-German Supercomputing Alliance (HLRN) under project bep00068.
\end{acknowledgments}

\bibliography{bibliography.bib}

\end{document}


\title{Time-dependent friction effects on vibrational infrared frequencies and line shapes of liquid water}


\author{Florian N. Br\"unig}
\affiliation{Freie Universit\"at Berlin, Germany}

\author{Otto Geburtig}
\affiliation{Freie Universit\"at Berlin, Germany}

\author{Alexander von Canal}
\affiliation{Freie Universit\"at Berlin, Germany}

\author{Julian Kappler}
\affiliation{Freie Universit\"at Berlin, Germany}

\author{Roland R. Netz}
\email{rnetz@physik.fu-berlin.de}
\affiliation{Freie Universit\"at Berlin, Germany}



\maketitle


\begin{acronym}[Bash]
 \acro{aiMD}{ab initio Molecular Dynamics}
 \acro{ATR}{attenuated total reflection}
 \acro{DFT}{density functional theory}
 \acro{FPT}{first-passage time}
 \acro{GLE}{generalized Langevin equation}
 \acro{GH}{Grote-Hynes}
 \acro{FTIR}{Fourier-transform infrared spectroscopy}
 \acro{FWHM}{full width at half maximum}
 \acro{IR}{infrared}
 \acro{LE}{Langevin equation}
 \acro{MD}{molecular dynamics}
 \acro{MFPT}{mean first-passage time}
 \acro{MFP}{mean first-passage}
 \acro{MSD}{mean squared displacement}
 \acro{NM}{normal-mode}
 \acro{PTP}{$p(\text{TP}|q)$}
 \acro{PME}{particle mesh Ewald~\cite{pronk2013gromacs}}
 \acro{PMF}{potential of mean force}
 \acro{PGH}{Pollak-Grabert-Hanggi}
 \acro{RC}{reaction coordinate}
 \acro{RDF}{radial distribution function}
 \acro{RTT}{round-trip time}
 \acro{TP}{transition path}
\end{acronym}

\thispagestyle{empty}

\newpage



\section{Infrared power spectra from linear-response theory}
\label{linearResponseSection}

Assuming linear response of an observable $x(t)$ with respect to a force that couples to an observable $y(t)$, the response function $\chi_{xy}(t)$ is related to the correlation function $C_{xy}(t')=\langle x (t+t') y(t)\rangle$ for $t\geq 0$ \cite{Kohler1972}
\begin{align}
\label{linearResponseSI}
\chi_{xy}(t) = - \frac{1}{ k_BT } \frac{d}{dt} C_{xy}(t),
\end{align}
where $k_BT$ is the thermal energy. Realizing that $\chi(t)$ is single-sided, i.e. $\chi(t)=0$ for $t<0$, the Fourier transform is calculated as
\begin{align}
\widetilde \chi_{xy}(\omega) &= - \frac{1}{ k_BT } \int_{-\infty}^{\infty} dt\ e^{i\omega t} \frac{d}{dt} C_{xy}(t)\nonumber \\
\widetilde \chi_{xy}(\omega) &= - \frac{1}{ k_BT } \left( C_{xy}(0) - i \omega \int_0^{\infty} dt\ e^{i\omega t} C_{xy}(t) \right)\nonumber \\
\widetilde \chi_{xy}(\omega) &= - \frac{1}{ k_BT } \left( C_{xy}(0) - i \omega \widetilde C_{xy}^+(\omega) \right),
\label{linearResponseFTSI}
\end{align}
where the superscript $^+$ denotes a single-sided Fourier transform.
In case of $x=y$, $C_{xx}(t)$ is an autocorrelation function, which is real and symmetric, therefore it follows for the imaginary part of the response function in Fourier space
\begin{align}
\widetilde \chi_{xx}''(\omega) &= \frac{1}{ k_BT } \omega \operatorname{Re} ( \widetilde  C^+_{xx}(\omega) ) \\
\label{omegaDoublePrimeCXX}
 &= \frac{1}{ k_BT } \frac{\omega}{2} \widetilde C_{xx}(\omega).
\end{align}
When computing the power spectra of a stochastic process $x(t)$, limited to the time domain $[0,L_t]$, the Wiener-Khintchine theorem, eq.~\eqref{WienerKintchineFT} in section \ref{WienerKintchineSection}, can be used to express $\widetilde C_{xx}(\omega)$ in terms of $\tilde x(\omega)$, turning eq.~\eqref{omegaDoublePrimeCXX} into
\begin{align}
\label{omegaDoublePrimeXOmega}
\widetilde \chi_{xx}''(\omega) = 
\frac{\omega}{2 k_BT L_t } | \tilde x(\omega)|^2.
\end{align}

When computing the power spectra of the observable $x(t)$ from the ensemble average of equilibrium trajectories, a decomposition of $x(t)$ into two parts $x (t)=x_1 (t)+ x_2 (t)$ gives rise to three contributions in the total power spectrum
\begin{align}
\nonumber
\omega \widetilde \chi_{xx}''(\omega)
&= \frac{\omega^2}{2 k_BT} \left[ \widetilde C_1(\omega)+ \widetilde C_2(\omega)+ 2\widetilde C_{1,2}(\omega) \right]\\
&= \omega \left[\widetilde \chi''_1(\omega) + \widetilde \chi''_2(\omega) + \widetilde \chi''_{1,2}(\omega) \right],
\end{align}
where the cross-correlation contribution $\chi''_{1,2}(\omega)$ was defined such that it equals the difference spectrum
\begin{align}
\nonumber
\widetilde \chi''_\mathrm{diff}(\omega)&= \widetilde \chi_{xx}''-\widetilde \chi''_1(\omega) - \widetilde \chi''_2(\omega) = \widetilde \chi''_{1,2}(\omega) \\
&= \frac{\omega}{k_BT} \widetilde  C_{1,2}(\omega).
\end{align}
A positive cross-correlation spectrum hints to in-phase motion, a negative cross-correlation spectrum to out-of-phase motion of $x_1 (t+t')$ and $x_2 (t)$ at a given frequency.

In case of $x(t)$ being the polarization $\bm p(t)$ of the system, coupling to an external electric field $\bm E(t)$, the dimensionless dielectric susceptibility $\chi(t)$ is given by
\begin{align}
\widetilde \chi(\omega) =  \frac{1}{V\epsilon_0 D} \langle \widetilde \chi_{\bm p \bm p}(\omega) \rangle,
\end{align}
where $\epsilon_0$ is the vacuum permittivity, $V$ is the system volume and an average is performed over the $D$ dimensions of $\bm p$.

\section{Decomposition of water infrared spectra into single-molecular and collective components}
\label{collectivitySection}

\begin{figure*}[hb]
\centering
\begin{overpic}[width=0.49\textwidth]{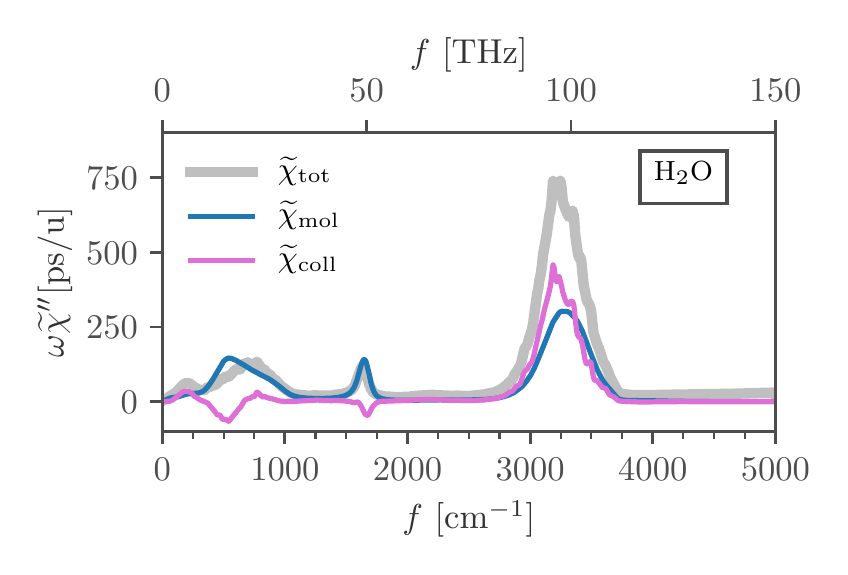}
\put(2,55){\large \bf A}
\end{overpic}
\begin{overpic}[width=0.49\textwidth]{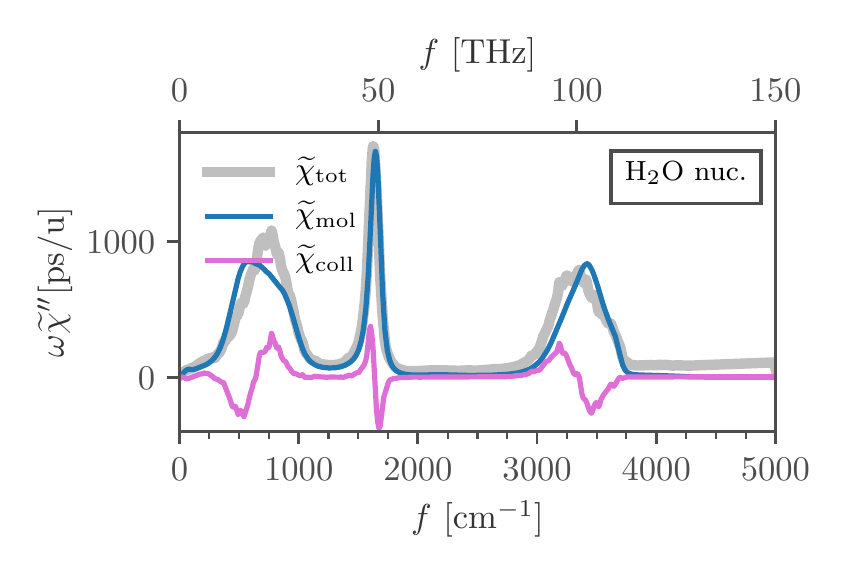}
\put(2,55){\large \bf B}
\end{overpic}
\caption{
Decomposition of the \ac{IR} spectra into total (grey solid lines), single-molecular (blue solid lines) and collective contributions (purple solid lines), obtained from the total dipole-moment trajectory from the \ac{aiMD} simulation of 256 H$_2$O molecules including nuclear and electronic charges after Wannier localization ({\bf A}) and an approximation using partial charges on the nuclear dynamics of the same \ac{aiMD} simulation ({\bf B}, see the following section \ref{atomCoordinatesSection} for details).
}
\label{h2o_pc}
\end{figure*}

Fig.~\ref{h2o_pc}A and B shows a complete decomposition of the total (nuclear and electronic) \ac{IR} spectra obtained from \ac{aiMD} simulations of 256 H$_2$O molecules (grey solid lines), as introduced in the main text, into single-molecular (blue solid lines) and collective components (purple solid lines), which follow from the molecular dipole moments $\bm p_i(t)$ of all the molecules in the bulk as \cite{Carlson2020}
\begin{align}
\widetilde \chi''_{\rm tot}(\omega) &= \widetilde \chi''_{\rm mol}(\omega) + \widetilde \chi''_{\rm coll}(\omega) \\
\sim \langle \sum_i \bm p_i(0) \sum_j \bm p_j(t) \rangle &= \sum_i \langle \bm p_i(0) \bm p_i(t) \rangle + \sum_i \langle \bm p_i(0)  \sum_{j \neq i} \bm p_j(t) \rangle.
\end{align}

The data clearly shows, that whereas in the OH stretching regime the collective effects are constructive and lead to an amplification of the total infrared spectrum with respect to the single-molecular spectrum, in the HOH bending regime the collective effects are destructive and lead to a slight decrease of the amplitude of the total \ac{IR} spectrum with respect to the single-molecular spectrum. In the regime of the librations between \SIrange{300}{800}{cm^{-1}} the collective effects contribute constructive as well as destructive. At the small signature around \SI{200}{cm^{-1}}, associated with translational vibrations of water molecules against each other, the collective effects contribute constructive.

\clearpage

\section{Infrared spectra from nuclear coordinates}
\label{atomCoordinatesSection}

\begin{figure*}[hb]
\centering
\begin{overpic}[width=0.5\textwidth]{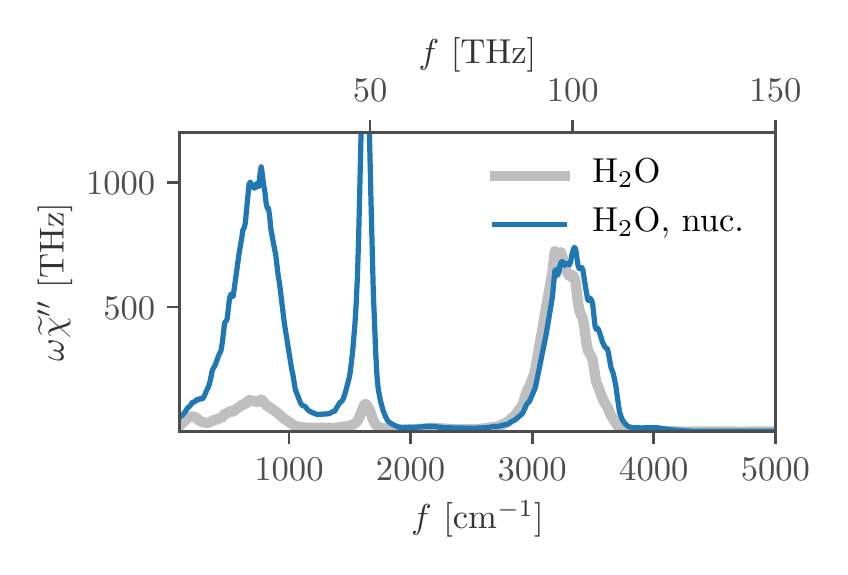}
\end{overpic}
\caption{Comparison of \ac{IR} spectra obtained from the total dipole-moment trajectory including nuclear and electronic charges from the \ac{aiMD} simulation (grey solid line) and an approximation using partial charges on the nuclear dynamics of the same \ac{aiMD} simulation (blue solid line).}
\label{h2o_classical}
\end{figure*}

Fig.~\ref{h2o_classical} shows a comparison of the \ac{IR} spectrum of the \ac{aiMD} simulation, using the total dipole moment including nuclear and electronic charges and a nuclei-only power spectrum, which is obtained by assigning the partial charges $\SI{1}{e}$ to the hydrogen atoms and $\SI{-2}{e}$ to the oxygen atoms of the \ac{aiMD} simulation in post processing. Both spectra are qualitatively similar but the nuclei-only spectrum is significantly increased at lower frequencies, due to the neglect of electronic polarization effects. In the OH-stretching regime at around \SI{3300}{cm^{-1}} both spectra coincidentally have a similar amplitude. Interestingly the nuclei-only spectrum is blue-shifted with respect to the spectrum from nuclear and electronic charges, indicating a slow down of the total dipole-moment dynamics in this regime due to the electronic degrees of freedom.

\clearpage 
\section{Extraction and parametrization of the potentials}
\label{FreeEnergyFitSection}

\begin{figure*}[htb]
\centering
\begin{overpic}[width=0.49\textwidth]{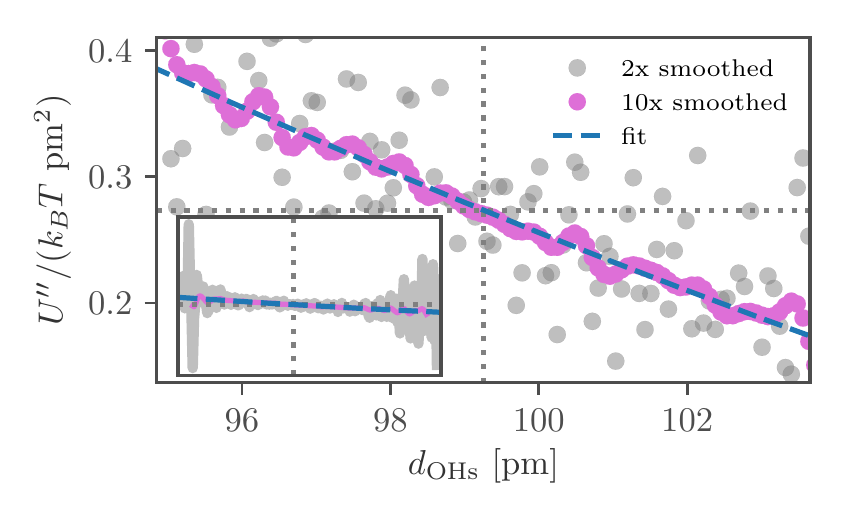}
\put(2,55){\large \bf A}
\end{overpic}
\begin{overpic}[width=0.49\textwidth]{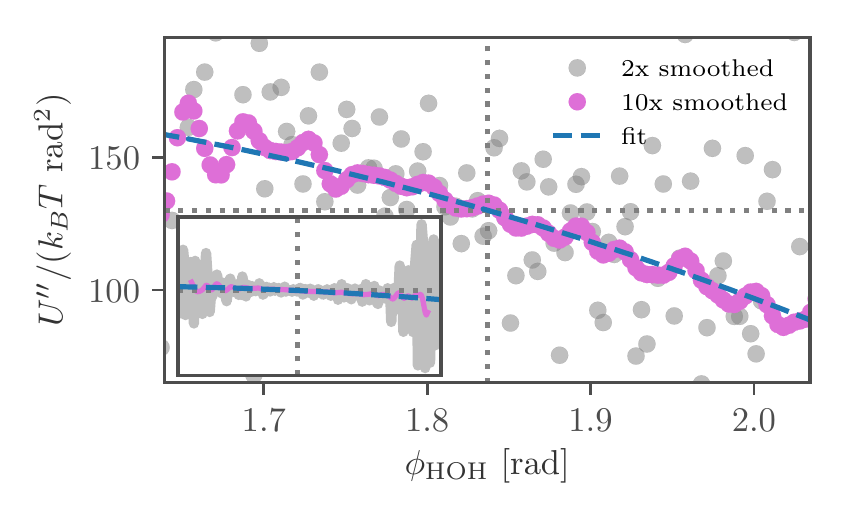}
\put(2,55){\large \bf B}
\end{overpic}
\begin{overpic}[width=0.49\textwidth]{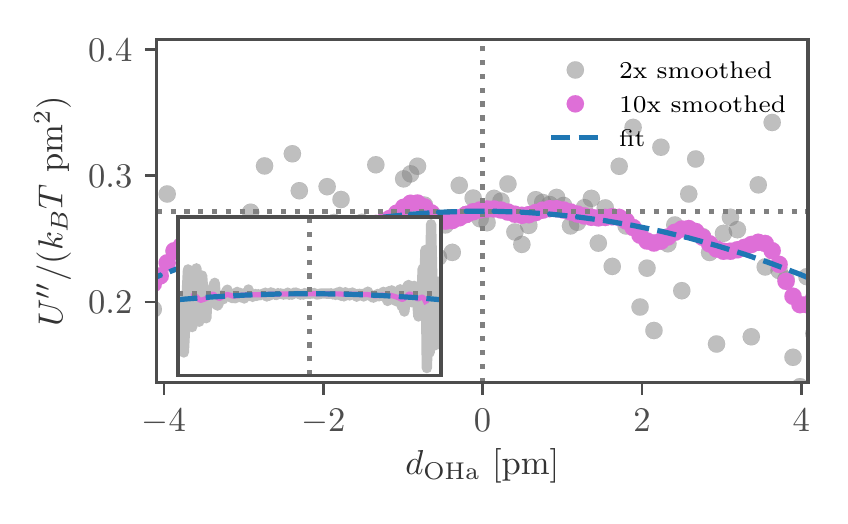}
\put(2,55){\large \bf C}
\end{overpic}
\begin{overpic}[width=0.49\textwidth]{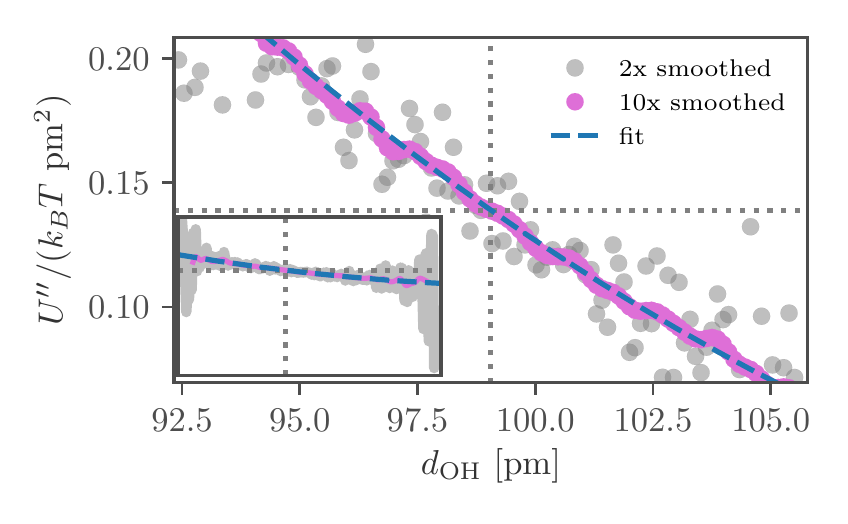}
\put(2,55){\large \bf D}
\end{overpic}
\caption{Second derivatives of the potential, obtained numerically and iteratively smoothed (grey and purple dots), of the various vibrational coordinates ({\bf A:} $\dohs$ coordinate, {\bf B:} $\hoh$ coordinate, {\bf C:} $\doha$ coordinate and {\bf D:} $\doh$ coordinate.). The second derivatives of the fit functions are shown as blue broken lines. The vertical grey dotted lines denote the position of the minima at $x_0$ and the horizontal grey dotted lines denote the values of the harmonic fit parameters $k$. The insets show a wider range of data, for which $U(x_i)<8\,k_BT$.}
\label{ddU}
\end{figure*}

From the entire set of trajectories of a vibrational coordinate a histogram is created with 50 equidistant bins centered at $x_i$. From this histogram $h(x_i)$ the potential (or free energy) is calculated as
\begin{align}
U(x_i) = -\log(h(x_i)) k_BT,
\end{align}
where the potential if shifted vertically so that $\min(U[x_i])=0$. The largest and smallest value of $x_i$ where $U(x_i) < 8\,k_BT$ are the edges for a new histogram with 99 equidistant bins. From this histogram the potential is again calculated as shown above. This potential is then fitted to a 4th-order polynom, $U(x)=a+bx+cx^2+dx^3+ex^4$, using the Levenberg-Marquardt algorithm. The position of the minimum of the fit function,  $x_0$, is subsequently obtained using Newton's method, both implemented in scipy v1.5. Eventually, $U(x)$ is rewritten as
\begin{align}
U(x) = & k_0 + \frac{k}{2}(x-x_0)^2 + \frac{k_3}{3}(x-x_0)^3 + \frac{k_4}{4}(x-x_0)^4\\
k_4 = & 4e\\
k_3 = & 3(d + k_4x_0)\\
k = & 2(c + k_3x_0 - \frac{6}{4}k_4x_0^2)\\
k_0 = & a - \frac{k_2}{2}x_0^2+\frac{k_3}{3}x_0^3-\frac{k_4}{4}x_0^4.
\end{align}
Note that $k_0$ is small but can be non-zero because of discretization effects. 
The results of the fits for the different vibrational coordinates are presented in fig.~\ref{ddU}A--D, by comparing the second derivatives of the fit functions, $U''(x)=k+2k_3(x-x_0)+3k_4(x-x_0)^2$, (blue broken lines) with second derivatives obtained numerically from the data, $U''(x_i) = (U(x_{i+1})-2U(x_{i+1})+U(x_{i-1}))/(x_i-x_{i-1})^2$, using histograms with 300 equidistant bins. The numerically obtained derivates are smoothed (grey and purple dots) by iterative convolution with a flat window function, which has a width of three bins. The plots illustrate that for all vibrational coordinates the fit functions and particularly the harmonic fit parameter $k$ (shown as horizontal grey dotted lines) match well the numerically obtained curvature of the potential around the position of the minima at $x_0$ (shown as vertical grey dotted lines).

\section{Extraction of the time-dependent friction kernels}
\label{kernelExtractionSection}
The extraction of time-dependent friction kernels from trajectories with an anharmonic system potential is performed using a modification of a recently published approach \cite{Daldrop2018b} and based on previous work by \citet{Harp1970}. The derivation starts from the \ac{GLE}
\begin{align}
\label{eq:gleah}
m \ddot{x}(t) = - \int_0^{t} \Gamma(t-t') \dot x(t') dt' - \nabla U[x(t)] + F_R(t),
\end{align}
which is multiplied by the initial velocity $\dot{x}(0)$ and ensemble averaged
\begin{align}
\label{eq:gle_volterra_1st}
\begin{split}
m\left<\dot{x}(0)\ddot{x}(t)\right>=
-\int_0^{t} dt'\, \Gamma(t') &\left<\dot{x}(0)\dot{x}(t-t')\right>-\left<\dot{x}(0)\nabla U[x(t)]\right>,
\end{split}
\end{align}
assuming the random force $F_R(t)$ to be uncorrelated with $\dot{x}(0)$. Defining the correlation functions as
\begin{align}
C_{vv}(t)&=\left<\dot{x}(0)\dot{x}( t)\right>,\\
C_{v\nabla U}(t)&=\left<\dot{x}(0)\nabla U[x( t)]\right>,\\
C_{x\nabla U}(t)&=\left<x(0)\nabla U[x( t)]\right>,
\end{align}
eq.~\eqref{eq:gle_volterra_1st} is written as
\begin{align}
m \frac{d}{dt} C_{vv}(t) =
-\int_0^{t} dt'\, \Gamma(t') & C_{vv}(t-t')-C_{v\nabla U}(t),
\end{align}
and integrated over time
\begin{align}
\label{eq:gle_volterra_1st_Int}
m C_{vv}(t) - m C_{vv}(0) =
-\int_0^{t} dt''\int_0^{t''} dt'\, \Gamma(t') & C_{vv}(t''-t')- \int_0^{t} dt''\  C_{v\nabla U}(t'').
\end{align}
For the terms on the right hand side, one finds
\begin{align}
\int_0^{t} dt''\,  C_{v\nabla U}(t'') &= \int_0^{t} dt''\, \left<\dot{x}(0)\nabla U[x( t'')]\right> \nonumber \\
&= \int_0^{t} dt''\, \left<\dot{x}(-t'')\nabla U[x(0)]\right> \nonumber \\
&= \left<\nabla U[x(0)] \int_0^{t} dt''\,  \dot{x}(-t'')\right> \nonumber \\
&= \left<\nabla U[x(0)] \int_{-t}^{0} dt''\,  \dot{x}(t'')\right> \nonumber \\
&= \left<\nabla U[x(0)] x(0)\right> - \left<\nabla U[x(0)] x(-t)\right>  \nonumber \\
\label{eq:RS_UX}
&= C_{x\nabla U}(0) - C_{x\nabla U}(t),
\end{align}
and
\begin{align}
\int_0^{t} dt''\int_0^{t''} dt'\, \Gamma(t') C_{vv}(t''-t') &=  \int_0^{t} dt''\int_0^{t''} dt'''\, \Gamma(t''-t''') C_{vv}(t''') \nonumber \\
&=  \int_0^{t} dt''\int_0^{t''} dt'''\, \Gamma(t''-t''') C_{vv}(t''') \nonumber \\
&=  \int_0^{t} dt'''\, C_{vv}(t''') \int_{t'''}^{t} dt''\, \Gamma(t''-t''') \nonumber \\
&=  \int_0^{t} dt'''\, C_{vv}(t''') \int_{0}^{t-t'''} dt''\, \Gamma(t'') \nonumber \\
\label{eq:RS_CG}
&=  \int_0^{t} dt'''\, C_{vv}(t''') G(t-t'''),
\end{align}
where
\begin{align}
G(t) = \int_0^t dt'\, \Gamma(t'),
\end{align}
is the integral over the friction kernel. Inserting eq.~\eqref{eq:RS_UX} and eq.~\eqref{eq:RS_CG} into eq.~\eqref{eq:gle_volterra_1st_Int}, one obtains
\begin{align}
\label{eq:gle_volterra_G_preFinal}
m C_{vv}(t) - m C_{vv}(0) = - \int_0^{t} dt'\, C_{vv}(t') G(t-t') - C_{x\nabla U}(0) + C_{x\nabla U}(t).
\end{align}
The mass is found to be 
\begin{align}
\label{eq:gle_volterra_G_mass}
m = \frac{C_{x\nabla U}(0)}{C_{vv}(0)},
\end{align}
which follows from multiplying the \ac{GLE} eq.~\eqref{eq:gleah} by the initial position $x(0)$, averaging over the ensemble and evaluating at $t=0$
\begin{align}
m\left<x(0)\ddot{x}(0)\right>&= -\left<x(0)\nabla U[x(0)]\right> \\
\quad -m\left<\dot{x}(0)\dot{x}(0)\right>&= -\left<x(0)\nabla U[x(0)]\right>.
\end{align}
Using eq.~\eqref{eq:gle_volterra_G_mass} in eq.~\eqref{eq:gle_volterra_G_preFinal}, one finds
\begin{align}
\label{eq:gle_volterra_G}
 C_{vv}(t) \frac{C_{x\nabla U}(0)}{C_{vv}(0)} = - \int_0^{t} dt'\, C_{vv}(t') G(t-t') + C_{x\nabla U}(t).
\end{align}

Eq.~\eqref{eq:gle_volterra_G} is a Volterra equation of first kind which can be discretized in time, $t=i\Delta t$, and solved numerically,
\begin{align}
C_{vv}^i \frac{C_{x\nabla U}^0}{C_{vv}^0} &=-\sum_{j=0}^{i}w_{j,i}\Delta t \, C_{vv}^{i}G^{i-j} +C_{x\nabla U}^i\\
&=-G^i w_{i,i}\Delta t \, C_{vv}^{0}-\sum_{j=0}^{i-1}w_{j,i}\Delta t \, C_{vv}^{i}G^{i-j} +C_{x\nabla U}^i,
\end{align}
where $w_{j,i}$ are integration weights and the corresponding iteration relation reads
\begin{align}
\label{eq:kernel_global_vlt1}
G^i=-\frac{1}{\omega_{i,i} \Delta t C_{vv}^0}\left( \sum_{j=0}^{i-1} \omega_{i,j}\Delta t C_{vv}^{i}G^{i-j} + \frac{C_{x\nabla U}^0}{C_{vv}^0} C_{vv}^i-C_{x\nabla U}^i \right).
\end{align}
Thus by calculating the necessary correlation functions, $C_{vv}^i=C_{vv}(i\Delta t)$ and $C_{x\nabla U}^i = C_{x\nabla U}(i\Delta t)$, from equilibrium trajectories, the integral of the time-dependent friction kernel, $G^i=G(i\Delta t)$, can be obtained by iteration from the initial value $G^0=0$. The friction kernel $\Gamma^i=\Gamma(i\Delta t)$ is subsequently obtained by a numerical derivative.

\section{Power spectrum of the Generalized Langevin Equation}
\label{gleResponseSection}

A random process $x(t)$ with mass $m$ and subject to time-dependent (or frequency-dependent) friction $\Gamma(t)$ in a harmonic potential $U(x)=\frac{k}{2}x^2$ can be described by the \acl{GLE}
\begin{align}
\label{eq:gle}
m\ddot{x}(t)=-\int_0^{t} dt'\, \Gamma(t-t')\dot{x}(t') -k x(t)+F_R(t),
\end{align}
obeying for the random force $F_R(t)$ the fluctuation-dissipation relation $\langle F_R(t) F_R(t') \rangle = k_BT \Gamma(t-t')$. A Fourier transform gives
\begin{align}
\label{eq:gleFT}
- \omega^2 m\tilde {x}(\omega)=i \omega \widetilde \Gamma^+(\omega)\tilde {x}(\omega) - k \tilde x(\omega)+\widetilde F_R(\omega),
\end{align}
as well as
\begin{align}
\langle \widetilde F_R(\omega) \widetilde F_R(\omega') \rangle &= k_BT \int_{-\infty}^{\infty} dt \ e^{i\omega t} \int_{-\infty}^{\infty} dt' \ e^{i\omega' t'}  \Gamma(t-t')\\
 &= k_BT \int_{-\infty}^{\infty} dt \ e^{i\omega t} \int_{-\infty}^{\infty} dt' \ e^{i\omega' t'}  \Gamma(t-t') \nonumber\\
 &= k_BT \int_{-\infty}^{\infty} dt' \ e^{i(\omega'+\omega) t'} \int_{-\infty}^{\infty} dt \ e^{i\omega (t-t')}  \Gamma(t-t') \nonumber \\
 \label{eq:noiseFT}
 &= 2\pi k_BT \delta(\omega'+\omega)  \widetilde \Gamma(\omega)
\end{align}
The absorbed power $\omega \widetilde \chi''(\omega)$ is derived in Fourier space from linear response to an external force $F_{\rm ext}$, which is calculated according to
\begin{align}
\widetilde \chi(\omega) &= \frac{\tilde x(\omega)}{\widetilde F_{\mathrm{ext}}(\omega)} \\
\label{eq:chiFT}
&= (k-m\omega^2-i\widetilde \Gamma^+(\omega)\omega)^{-1},
\end{align}
and using eqs.~\eqref{eq:noiseFT} and \eqref{eq:chiFT} the autocorrelation in Fourier space is given by
\begin{align}
\widetilde C_{xx}(\omega) &= \int_{-\infty}^{\infty} dt \ e^{i\omega t} \langle x(t) x(0) \rangle \\
&= \frac{1}{4\pi^2} \int_{-\infty}^{\infty} dt \ e^{i\omega t}  \int_{-\infty}^{\infty} d\omega' \ e^{-i\omega' t} \int_{-\infty}^{\infty} d\omega'' e^{-i\omega'' 0} \langle   \tilde x(\omega') \tilde x(\omega'') \rangle  \nonumber \\
&= \frac{1}{4\pi^2} \int_{-\infty}^{\infty} d\omega' \ 2 \pi \delta(\omega-\omega') \int_{-\infty}^{\infty} d\omega'' \langle   \tilde x(\omega') \tilde x(\omega'') \rangle  \nonumber \\
&= \frac{1}{2\pi} \int_{-\infty}^{\infty} d\omega'' \langle   \tilde x(\omega) \tilde x(\omega'') \rangle \\
&= \frac{1}{2\pi} \int_{-\infty}^{\infty} d\omega'' \langle   \widetilde \chi(\omega) \widetilde F_R(\omega) \widetilde \chi(\omega'')\widetilde F_R(\omega'')\rangle  \nonumber \\
&= k_BT \int_{-\infty}^{\infty} d\omega'' \widetilde \chi(\omega) \widetilde \chi(\omega'') \delta(\omega''+\omega)  \widetilde \Gamma(\omega)  \nonumber \\
&= k_BT \widetilde \chi(\omega) \widetilde \chi(-\omega) \widetilde \Gamma(\omega),
\end{align}
which is further simplified by realizing that the Fourier transform of the purely real function $\chi(t)$ is even $\widetilde \chi(-\omega) = \widetilde \chi^*(\omega)$
\begin{align}
\widetilde C_{xx}(\omega) &= k_BT \widetilde \chi(\omega) \widetilde \chi^*(\omega) \widetilde \Gamma(\omega)  \nonumber \\
&= k_BT \left| \widetilde \chi(\omega)\right|^2 \widetilde \Gamma(\omega)  \nonumber \\
&= \frac{k_BT \widetilde \Gamma(\omega)}{\left|k-m\omega^2-i\widetilde \Gamma^+(\omega)\omega\right|^2}.
\end{align}
From eq.~\eqref{omegaDoublePrimeCXX} the power spectrum follows
\begin{align}
\omega \widetilde \chi''(\omega)  =  \frac{\omega^2\ \text{Re} \widetilde \Gamma^+(\omega)}{\left|k-m\omega^2-i\widetilde \Gamma^+(\omega)\omega\right|^2}.
\end{align}

\section{Power spectrum of the damped harmonic oscillator}
\label{hoResponseSection}

The absorbed power $\omega \widetilde \chi''(\omega)$ of the damped harmonic oscillator described by the differential equation
\begin{align}
m\ddot x(t) = -\gamma \dot x(t) - k x(t) + F_{\mathrm{ext}} (t),
\label{HO}
\end{align}
is computed from the linear response in Fourier space
\begin{align}
\widetilde \chi(\omega) &= \frac{\tilde x(\omega)}{\tilde F_{\mathrm{ext}}(\omega)} \\
&= (k-m\omega^2-i\gamma\omega)^{-1} \\
&= \frac{k-m\omega^2+i\gamma\omega}{(k-m\omega^2)^2+\gamma^2\omega^2},
\end{align}
where $\tilde x(\omega)$ is the oscillating variable, $m$ is the mass, $\gamma$ the friction coefficient, $k$ the spring constant of the harmonic potential and $\tilde F_{\mathrm{ext}}(\omega)$ an external force.
For the power spectrum follows
\begin{align}
\omega \widetilde \chi''(\omega)  = \frac{\gamma\omega^2}{(k-m\omega^2)^2+\gamma^2\omega^2},
\label{LinResHO}
\end{align}
which by introducing the time scales $\tau = 2 \gamma / k$, $\tau_{\omega} = \sqrt{m / k}$ and length scale $L$ with $L^2 = 2 k_BT/k$ converts to
\begin{align}
\label{LinResHORed}
\omega \widetilde \chi''(\omega)  = \frac{L^2}{k_BT}\frac{\tau \omega^2}{4(1-\tau_{\omega}^2 \omega^2)^2+\tau^2\omega^2}.
\end{align}
In spectroscopy this is known as a Lorentzian line shape \cite{Schrader1995}, which in the overdamped case, $\tau_{\omega} \to 0$, reads
\begin{align}
\label{OverdampedDebyeSpectrum}
\omega \widetilde \chi''(\omega)  &= \frac{\gamma\omega^2}{ k^2+\gamma^2\omega^2}\\
&= \frac{L^2}{ k_BT}\frac{\tau \omega^2}{4+\tau^2\omega^2}.
\end{align}
Eq.~\eqref{OverdampedDebyeSpectrum} is also known as the Debye line shape. 

\section{Parametrization of the extracted memory kernels}
\label{KernelFitSection}

The memory kernels $\Gamma(t)$, extracted from the \ac{aiMD} simulation data as described in section \ref{kernelExtractionSection}, are truncated at $\SI{50}{ps}$, over-sampled at a time resolution of $\Delta t = \SI{0.025}{fs}$ and subsequently Fourier transformed using the FFT algorithm implemented in numpy v1.19. The real part of the Fourier-transformed memory kernel is then fitted to a combination of $n$ exponential and $l$ oscillating memory kernels according to eq.~\eqref{eq:kernel_param} in the main text. The Fourier-transformed expressions of the fundamental kernels are found to be
\begin{align}
\widetilde{\Gamma}_{\rm osc}^+(\omega, a_i, \tau_i, \omega_i) &=\frac{a_i}{2}\Bigl( \frac{1+\frac{i}{\omega_i\tau_i}}{\frac{1}{\tau_i}+i\omega_i-i\omega}+ \frac{1-\frac{i}{\omega_i\tau_i}}{\frac{1}{\tau_i}-i\omega_i-i\omega}\Bigr) \\ 
\widetilde{\Gamma}_{\rm exp}^+(\omega, \gamma_i, \tau_i) &= \gamma_i \dfrac{1}{1-i\omega \tau_i}  \\
\tilde{\Gamma}^+(\omega)&= \sum\limits_{i=1}^n \widetilde{\Gamma}_{exp}^+(\omega, \gamma_i, \tau^e_i)+\sum\limits_{i=1}^{l}\widetilde{\Gamma}_{osc}^+(\omega, a_i, \tau^o_i, \omega_i)
\end{align}
The real part of the Fourier-transformed memory kernel $\widetilde{\Gamma}'^+(\omega)$ is fitted in the $(2n+3l)$-dimensional parameter space using the Levenberg-Marquardt algorithm implemented in scipy v1.5. $n$ and $l$ are iteratively increased until the fit quality does not improve significantly. The initial values for all $\gamma_i$, $\tau^e_i$, $a_i$,  $\tau^o_i$ and $\omega_i$ are chosen suitably. After fitting in Fourier space, another fit is performed in the time domain, by subtracting the $l$ oscillating fit functions from the memory kernels in the time domain and again fitting the $n$ exponential functions to the remainder. This significantly improves the fits for the long-time tails of the memory kernels.
The hereby obtained fit parameters for the friction memory kernels for the different vibrational coordinates in the bulk water system are summarized in tabs.~\ref{tab:ohs_kernelFitParams}--\ref{tab:oh_kernelFitParams}.

\begin{table}[hp]
\centering
\caption{Fit parameters according to eq.~\eqref{eq:kernel_param} in the main text of the friction memory kernel of the $\dohs$ coordinate.}
\label{tab:ohs_kernelFitParams}
\vspace{10pt}
\begin{tabular}{c c c}
$\gamma_i$ [u/ps]& $\tau_i^e$ [ps]& \\ \hline 
2758 & 0.115 &  \\ 
25941 & 1.28 &  \\ 
54005 & 5.72 &  \\ 
\hline $a_i$ [u/ps$^2$] & $\tau_i^o$ [ps] & $\omega_i$ [THz]\\ \hline 
8468 & 0.00944 & 170 \\ 
3525 & 0.0146 & 1290 \\ 
5985 & 0.0256 & 614 \\ 
1095 & 0.0599 & 675 \\ 
10560 & 0.0807 & 28.7 \\ 
6318 & 0.122 & 313 \\ 

\end{tabular}
\end{table}

\begin{table}[hp]
\centering
\caption{Fit parameters according to eq.~\eqref{eq:kernel_param} in the main text of the friction memory kernel of the $\hoh$ coordinate.}
\label{tab:hoh_kernelFitParams}
\vspace{10pt}
\begin{tabular}{c c c}
$\gamma_i$ [u/ps]& $\tau_i^e$ [ps]& \\ \hline 
661 & 0.582 &  \\ 
5210 & 3.08 &  \\ 
\hline $a_i$ [u/ps$^2$] & $\tau_i^o$ [ps] & $\omega_i$ [THz]\\ \hline 
322 & 0.0257 & 1280 \\ 
2139 & 0.0283 & 28.9 \\ 
2577 & 0.0322 & 953 \\ 
 37 & 0.0323 & 1590 \\ 
786 & 0.0394 & 110 \\ 
4600 & 0.0396 & 643 \\ 
374 & 0.0417 & 162 \\ 
435 & 0.0465 & 136 \\ 
4089 & 0.0532 & 612 \\ 
 69 & 0.056 & 252 \\ 
471 & 0.0612 & 84.4 \\ 
 49 & 0.0839 & 284 \\ 
 40 & 0.12 & 311 \\ 
2836 & 0.143 & 14.4 \\ 

\end{tabular}
\end{table}

\begin{table}[hp]
\centering
\caption{Fit parameters according to eq.~\eqref{eq:kernel_param} in the main text of the friction memory kernel of the $\doha$ coordinate.}
\label{tab:oha_kernelFitParams}
\vspace{10pt}
\begin{tabular}{c c c}
$\gamma_i$ [u/ps]& $\tau_i^e$ [ps]& \\ \hline 
5653 & 0.183 &  \\ 
32596 & 2.03 &  \\ 
\hline $a_i$ [u/ps$^2$] & $\tau_i^o$ [ps] & $\omega_i$ [THz]\\ \hline 
7584 & 0.0165 & 1280 \\ 
3186 & 0.021 & 300 \\ 
5051 & 0.0266 & 602 \\ 
3208 & 0.0361 & 141 \\ 
1320 & 0.0401 & 413 \\ 
2810 & 0.0415 & 644 \\ 
39665 & 0.0752 & 36.5 \\ 
1353 & 0.0768 & 677 \\ 
2010 & 0.0856 & 100 \\ 

\end{tabular}
\end{table}

\begin{table}[hp]
\centering
\caption{Fit parameters according to eq.~\eqref{eq:kernel_param} in the main text of the friction memory kernel of the $\doh$ coordinate.}
\label{tab:oh_kernelFitParams}
\vspace{10pt}
\begin{tabular}{c c c}
$\gamma_i$ [u/ps]& $\tau_i^e$ [ps]& \\ \hline 
1212 & 0.109 &  \\ 
9005 & 1.06 &  \\ 
19399 & 4.52 &  \\ 
\hline $a_i$ [u/ps$^2$] & $\tau_i^o$ [ps] & $\omega_i$ [THz]\\ \hline 
1343 & 0.0174 & 249 \\ 
1669 & 0.0264 & 616 \\ 
1544 & 0.0301 & 145 \\ 
420 & 0.0364 & 411 \\ 
137 & 0.0688 & 673 \\ 
845 & 0.0699 & 100 \\ 
10641 & 0.081 & 35.5 \\ 
1802 & 0.115 & 312 \\ 

\end{tabular}
\end{table}

\clearpage

\section{Simulation of the generalized Langevin equation}
\label{GLESimulationSection}

The \ac{GLE}
\begin{align}
m \ddot{x}(t) = - \int_0^{t} \Gamma(t-t') \dot x(t') dt' - \nabla U[x(t)] + F_R(t),
\end{align}
with a sum of $n$ exponential and $l$ oscillating memory kernels, analogous to eq.~\eqref{eq:kernel_param} in the main text,
\begin{align}
\Gamma(t)&= \sum\limits_{i=1}^n a_i\ e^{-t/\tau_i} + \sum\limits_{i=1}^{l} a_i\ e^{-t/\tau_i} \Bigl(\cos(\omega_i t)+\dfrac{1}{\tau_i\omega_i}\sin(\omega_i t )\Bigr),
\end{align}
can be efficiently simulated using a Markovian embedding (which for the oscillating components is derived in detail in section \ref{oscMemSection}) 
\begin{align}
\dot{x}(t) &= v(t) \\
m_x \dot{v}(t) &= - \nabla U[x(t)] + \sum\limits_{i=1}^n a_i [y_i(t)-x(t)]+ \sum\limits_{i=1}^{l} a_i[z_i(t)-x(t)]\label{eq:equation_of_motion_x} &\\
\gamma_i \dot{y}_i(t) &= a_i[x(t)-y_i(t)]+F_i(t) & \textrm{for $n$ exp. components} \label{eq:equation_of_motion_y} \\
\dot{z}_i(t) &= w_i(t) \\ 
m_i \dot{w}_i(t) &= -\gamma_i w_i(t)+a_i[x(t)-z_i(t)]+F_i(t)
& \textrm{for $l$ osc. components} \label{eq:equation_of_motion_z} 
\end{align}
For the $n$ exponential components we obtain $a_i = \frac{\gamma_i}{\tau_i^e}$ from the fit parameters $\gamma_i$ and $\tau_i^e$. To get $m_i$ and $\gamma_i$ for the $l$ oscillating components from the fit parameters $a_i$, $\tau^o_i$ and $\omega_i$ we use
\begin{align}
m_i &= a_i ((\tau^o_i)^{-2}+\omega_i^2)^{-1} \\
\gamma_i &= 2 \frac{m_i}{\tau^o_i}
\end{align}
The random force $F_i$ present in eqs.~\eqref{eq:equation_of_motion_y} and \eqref{eq:equation_of_motion_z} is
\begin{align}
F_i &= \sqrt{2k_BT \gamma_i  \delta t^{-1}} \Xi
\end{align}
Where $\Xi$ is a Gaussian random distribution  with zero mean and a standard deviation of one. In the numerical simulation we use the numpy.random.normal function to generate these random numbers. Eqs.~\eqref{eq:equation_of_motion_x}--\eqref{eq:equation_of_motion_z} are numerically solved using a 4th-order Runge-Kutta scheme to get the trajectory of $x$. The timestep of the simulation is \SI{1e-6}{\pico\second} and only every 500th value is stored. Thus the trajectory is evaluated at a time step of \SI{0.5}{fs}. The trajectory contains \num{2.5e7} values which corresponds to a simulation time of \SI{1.25e4}{\pico\second}. Each trajectory is divided into 50 parts of \SI{250}{\pico\second} and their spectra are calculated using the Wiener-Kintchine relation introduced in section \ref{WienerKintchineSection}. All 50 spectra are then averaged to give the final spectrum. Starting velocity and position were set to be zero.

The memoryless Langevin equation, with a friction constant $\gamma$ replacing the integral over the memory kernel
\begin{align}
m \ddot{x}(t) = - \gamma \dot x(t') - \nabla U[x(t)] + F_R(t),
\end{align}
is likewise simulated using a 4th-order Runge-Kutta scheme. The timestep of these simulations is \SI{0.1}{fs} for a total of $10^7$ steps, which corresponds to a simulation time of \SI{1000}{\pico\second}. Each trajectory is divided into 200 parts of \SI{5}{\pico\second} and their spectra are calculated using the Wiener-Kintchine relation introduced in section \ref{WienerKintchineSection}. All 200 spectra are then averaged to give the final spectrum.

\section{Oscillating memory kernels}
\label{oscMemSection}

\subsection{Oscillating memory kernels from Hamiltonian}

Oscillating memory kernels arise from inertial dynamics in orthogonal degrees of freedom at time scales of the order of the primary reaction coordinate. The relation of the time scales of the orthogonal coordinate and the memory kernel is understood from an explicit derivation of the dynamics of the primary coordinate $x$ with velocity $v$, linearly coupled to an orthogonal coordinate $y$ with velocity $w$ and coupling constant $k$. The Hamiltonian of the system is given as
\begin{align}
H = \frac{m}{2}v^2 + \frac{m_y}{2}w^2 + \frac{k}{2} (x-y)^2,
\end{align}
defining the Hamilton equations
\begin{align}
\dot x (t) &= v \\
m \dot v (t) &= k \left[ x(t)-y(t) \right]\\
\dot y (t) &= w (t)\\
m_y \dot w (t) &= - k \left[ x(t)-y(t) \right].
\end{align}

These Newtonian equations of motion can straight-forwardly be extended to Langevin equations by introducing random forces $F_x(t), F_y(t)$ and friction terms with coefficients $\gamma_x, \gamma_y$
\begin{align}
\label{ddx}
\dot x (t) &= v \\
\label{ddv}
m \dot v (t) &= -\gamma_x v(t) + k \left[ x(t)-y(t) \right] + F_x(t)\\
\label{ddy}
\dot y (t) &= w (t)\\
\label{ddw}
m_y \dot w (t) &= -\gamma_y w(t) - k \left[ x(t)-y(t) \right] + F_y(t).
\end{align}
The random forces are designed to introduce dynamics on the order of $k_BT$ but no net energy transfer into the system. Therefore they have zero mean, $\langle F_i(t)\rangle=0$, and their strength on the order of the friction forces $\langle F_i(t)F_i(t') \rangle=2 \gamma_i k_BT \delta(t-t')$, fulfilling the fluctuation-dissipation relation. 

To obtain the \ac{GLE} including the memory kernel we will now find a solution for $y$ using eq.~\eqref{ddw} and insert this solution in eq.~\eqref{ddv}. Eq.~\eqref{ddw} is essentially a driven damped harmonic oscillator, which can be solved in a general fashion using matrix exponentials and introducing the inertial time scale $\tau_{m_y} = m_y/\gamma_y$ and the memory time scale $\tau_{k_y} = \gamma_y/k$. First we write the coupled first order ordinary differential eqs.~\eqref{ddy} and \eqref{ddw} in the matrix form
\begin{align}
 \dot{\vec{y}}(t) &= A \vec{y}(t),
 \end{align}
 where
 \begin{align}
 \vec{y}(t) &=\begin{pmatrix} y \\ \dot{y}\end{pmatrix} \\
 A&=\begin{pmatrix} 0 & 1 \\ -(\tau_{m_y}\tau_{k_y})^{-1}  & - \tau_{m_y}^{-1} \end{pmatrix}, \\
A^{-1}&=\begin{pmatrix} - \tau_{m_y}^{-1} & (\tau_{m_y}\tau_{k_y})^{-1}  \\ -1 & 0\end{pmatrix}.
\end{align}

The solution is given as 
\begin{align}
\begin{split}
\vec{y}(t) &= \exp{\left(A[t-t_0]\right)} \vec{y}_0\\
&+ \int_{t_0}^t dt'\exp{\left(A[t-t']\right)} \begin{pmatrix}0\\-(\tau_{m_y}\tau_{k_y})^{-1} x(t') + m_y^{-1} F_y(t')\end{pmatrix},
\end{split} 
\end{align}
Next a partial integration over the integral containing $x(t)$ is performed
\begin{align}
\begin{split}
\vec{y}(t) &= \exp{\left(A[t-t_0]\right)} \vec{y}_0 \\
	   &- \left[ A^{-1} \exp{\left(A[t-t']\right)} \begin{pmatrix}0\\-(\tau_{m_y}\tau_{k_y})^{-1} x(t)\end{pmatrix} \right]_{t_0}^t \\
           &+ A^{-1} \int_{t_0}^t dt' \exp{\left(A[t-t']\right)} \begin{pmatrix}0\\-(\tau_{m_y}\tau_{k_y})^{-1} v(t') \end{pmatrix} \\
           &+ \int_{t_0}^t dt' \exp{\left(A[t-t']\right)} \begin{pmatrix}0\\ m_y^{-1} F_y(t') \end{pmatrix}.
\end{split}
\label{eq:ySolution}
\end{align}
The matrix exponential can be evaluated by diagonalizing the matrix $A$. After introducing $\omega_0 = \sqrt{(2\tau_{m_y})^{-2}-(\tau_{m_y}\tau_{k_y})^{-1}}$ the Eigenvalues $\lambda_{1,2}$ of $A$ are given as
\begin{align}
\lambda_{1,2} = -(2\tau_{m_y})^{-1} \pm \omega_0,
\label{eq:eigenvalues}
\end{align}
and the matrix exponential reads
\begin{align}
\begin{split}
\exp(At) &=\frac{1}{\lambda_2-\lambda_1} \begin{pmatrix} \lambda_2 e^{\lambda_1t}-\lambda_1 e^{\lambda_2t} & e^{\lambda_2t} - e^{\lambda_1t} \\ \lambda_1\lambda_2 \left( e^{\lambda_1t}- e^{\lambda_2t} \right) & \lambda_2 e^{\lambda_2t} - \lambda_1 e^{\lambda_1t} \end{pmatrix}
\end{split}\\
\begin{split}
&= e^{-t/(2\tau_{m_y})}  \begin{pmatrix} \cosh(\omega_0t) + \frac{\sinh(\omega_0 t)}{2\tau_{m_y}\omega_0}  & -\frac{\sinh(\omega_0t)}{ \omega_0}  \\ \frac{\sinh(\omega_0 t)}{\tau_{m_y}\tau_{k_y}\omega_0}  & \cosh(\omega_0 t) - \frac{\sinh(\omega_0 t)}{2\tau_{m_y}\omega_0}  \end{pmatrix},
\end{split}
\end{align}
giving
\begin{align}
\begin{split}
&A^{-1}\exp(At) =\tau_{m_y}\tau_{k_y} e^{-t/(2\tau_{m_y})} \\ 
&\begin{pmatrix} -\frac{1}{\tau_{m_y}} \cosh(\omega_0t) + \frac{1}{2\tau_{m_y}^2\omega_0} \sinh(\omega_0 t) +  \frac{1}{\tau_{m_y}\tau_{k_y}\omega_0} \sinh(\omega_0t)& -\frac{1}{2\tau_{m_y} \omega_0} \sinh(\omega_0t) - \cosh(\omega_0 t)\\ \frac{1}{\tau_{m_y}\tau_{k_y}} \cosh(\omega_0t) + \frac{1}{\tau_{m_y}\tau_{k_y}\omega_0} \sinh(\omega_0 t) & \frac{1}{\tau_{m_y}\tau_{k_y}\omega_0} \sinh(\omega_0 t) \end{pmatrix}.
\end{split}
\label{eq:mExpSolution}
\end{align}
The solution for $y(t)=\vec{y}_1(t)$ is now obtained from eqs.~\eqref{eq:ySolution}--\eqref{eq:mExpSolution} and by sending $t_0 \to -\infty$, i.e. assuming an equilibrated system
\begin{align}
\begin{split}
y(t) &= x(t) - \int_{-\infty}^tdt' v(t')e^{-(t-t')/(2\tau_{m_y})} \\
           & \left[ \cosh(\omega_0 (t-t')) + \frac{1}{2\tau_{m_y} \omega_0} \sinh(\omega_0(t-t')) \right] \\
           & - \int_{- \infty}^tdt' m_y^{-1} F_y(t') e^{-(t-t')/(2\tau_{m_y})} \frac{1}{ \omega_0} \sinh(\omega_0(t-t')).
\end{split}
\label{eq:yFinal}
\end{align}
To obtain the generalized Langevin equation for $x(t)$ containing the oscillating memory kernel
\begin{align}
\Gamma(t)= k e^{-t/(2\tau_{m_y})} \left[ \cosh(\omega_0t) + \frac{1}{2\tau_{m_y} \omega_0} \sinh(\omega_0t) \right],
\label{eq:oscKernel}
\end{align}
eq.~\eqref{eq:yFinal} is inserted in eq.~\eqref{ddv}
\begin{align}
\begin{split}
m\dot v(t) &= -\gamma_x v(t) +  F_x(t) + \int_{-\infty}^tdt' v(t') \Gamma(t-t') + F'_y(t).
\end{split}
\label{eq:gleOscFinal}
\end{align}

\subsection{Underdamped limit: oscillating kernel}
The kernel of eq.~\eqref{eq:oscKernel} can be analysed for two limiting cases. For Im$(\omega_0) > (2\tau_{m_y})^{-1}$, which is equivalent to $4\tau_{m_y} > \tau_{k_y}$, the underdamped case, $\omega_0$ is purely imaginary
and the kernel is
\begin{align}
\label{eq:oscKernelLimit}
\Gamma(t)= k e^{-t/(2\tau_{m_y})} \cos(\mathrm{Im}(\omega_0) t) + \frac{1}{2 \tau_{m_y}\mathrm{Im}(\omega_0) }\sin(\mathrm{Im}(\omega_0) t).
\end{align}

The integral is evaluated to
\begin{align}
\gamma=\int_0^{\infty}dt\Gamma(t) =k \frac{4\tau_{m_y}}{1 - 4\tau_{m_y}^2 \mathrm{Im}(\omega_0)^2}=k\tau_{k_y}=\gamma_y,
\label{eq:oscKernelIntUD}
\end{align}

\subsection{Overdamped limit: exponential kernel}

For Re$(\omega_0) < (4\tau_{m_y})^{-1}$ and $\tau_{k_y} > (4\tau_{m_y})$, the overdamped case, it follows that $\omega_0$ is purely real. The integral equally gives

\begin{align}
\gamma=\int_0^{\infty}dt\Gamma(t) =k \frac{4\tau_{m_y}}{1 - 4\tau_{m_y}^2 \omega_0^2}=k\tau_{k_y}=\gamma_y.
\label{eq:oscKernelIntOD}
\end{align}

\clearpage

\section{Decomposition of the OH stretch mode}
\label{ohDecompSection}

\begin{figure*}[htb]
\centering
\begin{overpic}[width=0.5\textwidth]{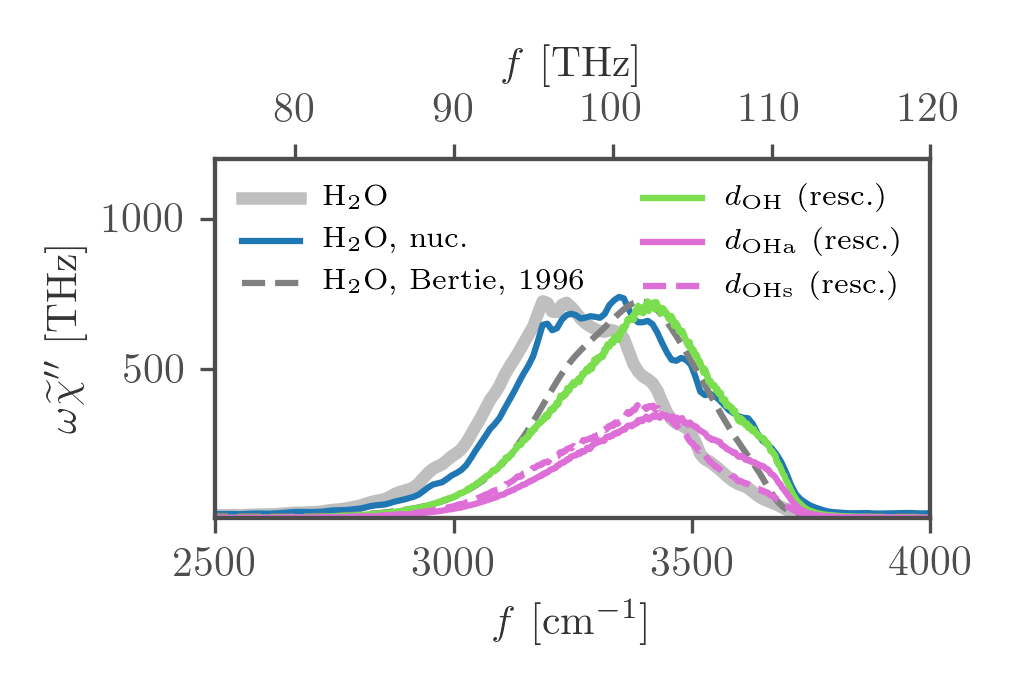}
\end{overpic}
\caption{Comparison of \ac{IR} spectra in the OH-stretching regime obtained from the total dipole-moment trajectory including nuclear and electronic charges from the \ac{aiMD} simulation of 256 H$_2$O molecules (grey solid line), an approximation using partial charges on the nuclear dynamics of the same \ac{aiMD} simulation (blue sold line, introduced in section \ref{atomCoordinatesSection}), experimental \ac{FTIR} data (grey broken line) \cite{Bertie1996} and rescaled power spectra of the $\doh$ (green solid line), $\dohs$ (purple solid line) and $\doha$ (purple broken line) vibrational coordinates.}
\label{oh_decomp}
\end{figure*}

As introduced in the main text, vibrational coordinates for the OH stretch mode are given by $\dohs = ({\doh}_1 + {\doh}_2)/2$ for the symmetric stretching-vibration and $\doha = ({\doh}_1 - {\doh}_2)/2$ for the anti-symmetric stretching vibration. Fig.~\ref{oh_decomp} compares the power spectra of these coordinates, obtained from averaging over all 256 H$_2$O molecules in the \ac{aiMD} simulation, to power spectra of the $\doh$ coordinate, experimental \ac{FTIR} data and the \ac{IR} spectra obtained from the total dipole-moment trajectory including nuclear and electronic charges from the \ac{aiMD} simulation and an approximation using partial charges on the nuclear dynamics of the \ac{aiMD} simulation, as introduced in section \ref{atomCoordinatesSection}. The power spectrum of the $\doh$ coordinate is blue-shifted with respect to the power spectra considering the total dipole moment, indicating a slow down of the collective nuclear and electronic dynamics with respect to the single-molecule dynamics, as discussed in section \ref{collectivitySection}. This could be similar to the effect of the slow down observed in the \ac{IR} spectra including electronic degrees of freedom compared to the nuclei-only \ac{IR} spectra, discussed in section \ref{atomCoordinatesSection}. As expected for the decomposition, the power spectra of $\dohs$ and $\doha$ exactly sum up to the $\doh$ power spectrum, there exists no remaining cross-correlation spectrum, so they form orthogonal coordinates.

The effective masses of the coordinates are determined by the equipartition theorem $m_{\mathrm{OH-sy.}} = k_BT /\langle \dot \dohs^2 \rangle=\SI{1.90104}{u}$, $m_{\mathrm{OH-as.}} = k_BT /\langle \dot \doha^2 \rangle=\SI{1.84571}{u}$ and $m_{\mathrm{OH}} = k_BT /\langle \dot \doh^2 \rangle=\SI{0.936477}{u}$. Analytic estimates are determined by the reduced mass of the $\doh$ coordinate $m_{\rm OH}=(m_{\rm O} m_{\rm H})/(m_{\rm O} + m_{\rm H}) = \SI{0.947617}{u}$. It follows for the masses of the $\dohs$ and $\doha$ coordinates $m_{\mathrm{OH-as.}} = m_{\mathrm{OH-sy.}}= 2 m_{\mathrm{OH}} = \SI{1.89523}{U}$.

\begin{figure*}[htp]
\centering
\begin{overpic}[width=0.49\textwidth]{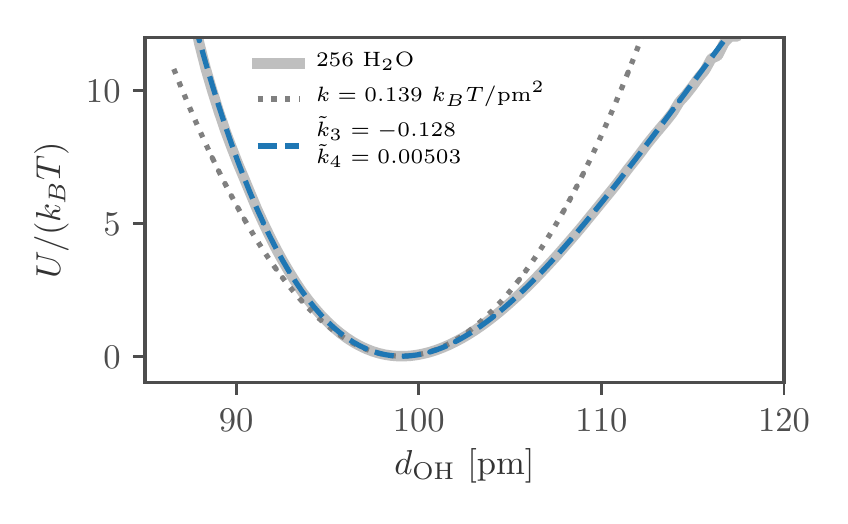}
\put(2,55){\large \bf A}
\end{overpic}
\begin{overpic}[width=0.49\textwidth]{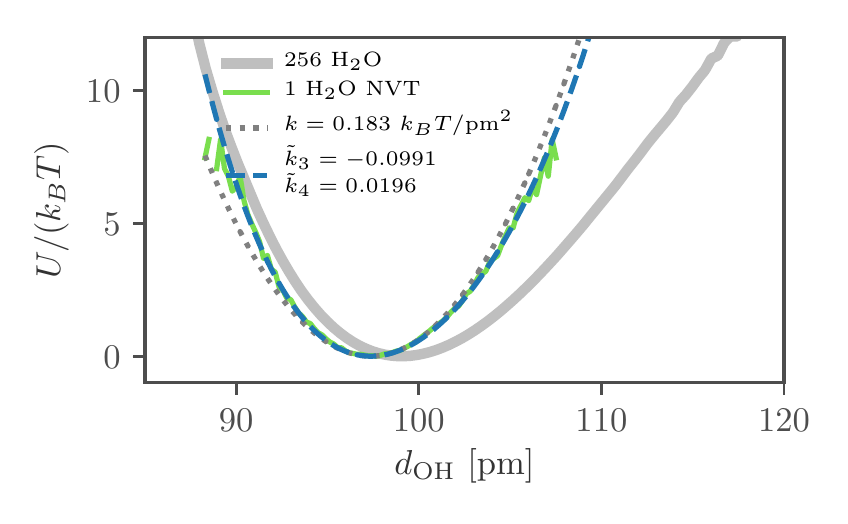}
\put(2,56){\large \bf B}
\end{overpic}
\begin{overpic}[width=0.49\textwidth]{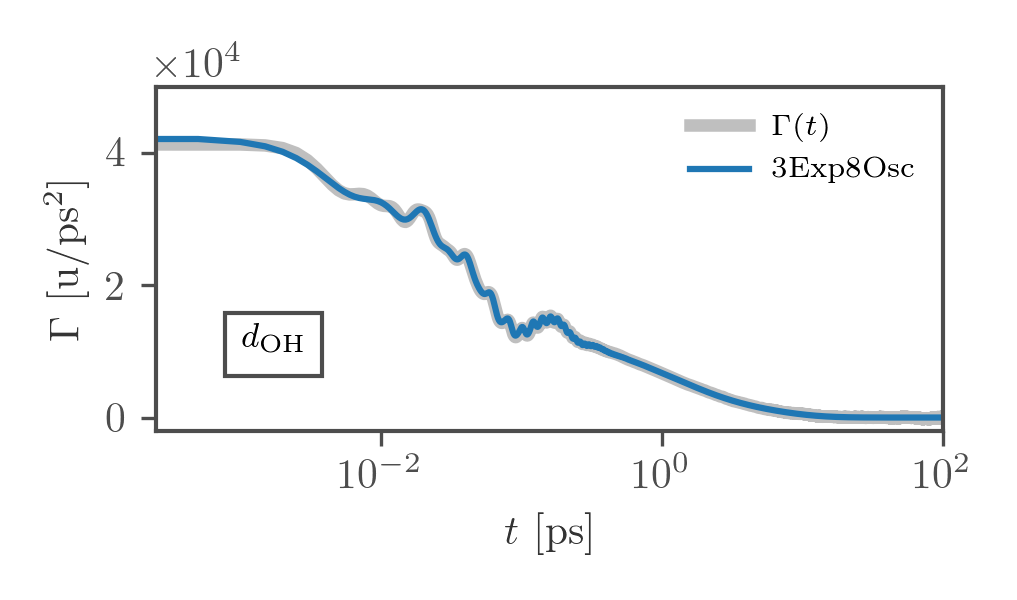}
\put(2,56){\large \bf C}
\end{overpic}
\begin{overpic}[width=0.49\textwidth]{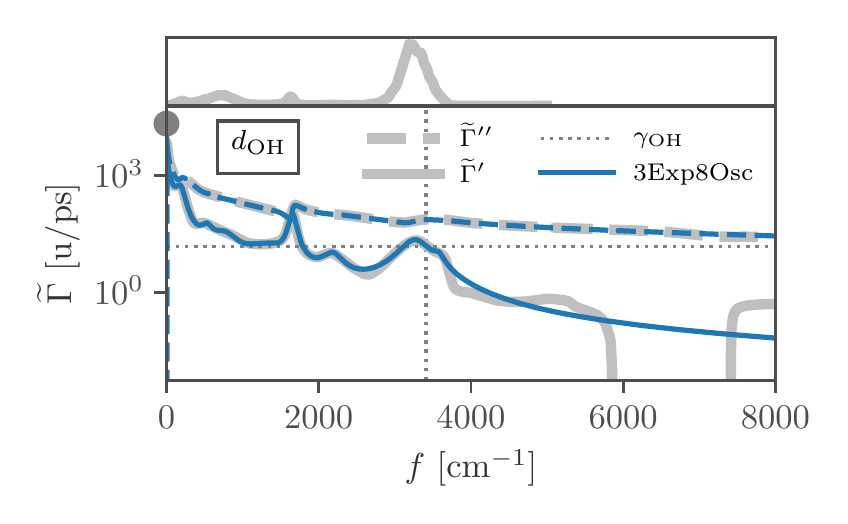}
\put(2,56){\large \bf D}
\end{overpic}
\begin{overpic}[width=0.49\textwidth]{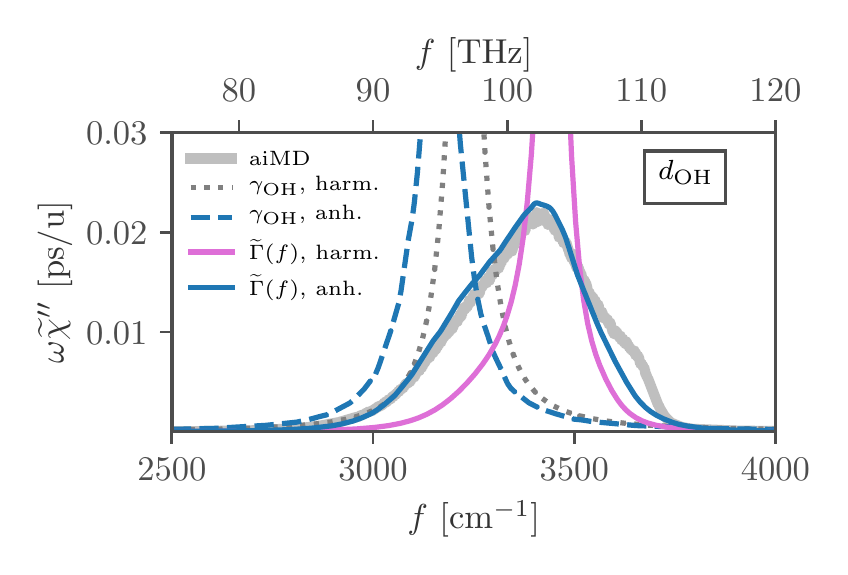}
\put(2,55){\large \bf E}
\end{overpic}
\caption{\footnotesize Results for the stretch coordinate  $\doh$  from aiMD simulations.
{\bf A, B:} Potential $U(\doh)$ for 256 H$_2$O in the liquid phase (grey solid line) and for  a single H$_2$O  (green solid line in B), both
compared to the non-harmonic fit according to eq.~\eqref{eq:pot_param} in the main text (blue broken line) and the harmonic part  (grey dotted line).
{\bf C, D:} Friction as a function of time and  frequency  (grey lines) compared with
the fit according to eq.~\eqref{eq:kernel_param} in the main text (blue lines). Real and imaginary parts in (D) are shown as solid and broken lines,
the spectrum on top is the full absorption spectrum from aiMD. The dotted horizontal line in (D) shows the constant  real friction $\gamma_{\rm OH}= \widetilde{\Gamma}'(f_{\rm OH})$ evaluated at the OH stretch vibrational frequency $f_{\rm OH}=\SI{3400}{cm^{-1}}$.
The grey circle denotes the static friction $\widetilde \Gamma'(0)$.
{\bf E:} Power spectrum $\omega \widetilde{\chi}''$
(grey solid line) compared to  models of varying complexity:
Lorentzian with harmonic potential and constant friction $\gamma_{\rm OH}$  (grey dotted line),
non-harmonic potential  and constant friction $\gamma_{\rm OHs}$  (blue broken line),
harmonic potential and frequency-dependent friction $\widetilde{\Gamma}(f)$ (purple solid line),
non-harmonic potential  and frequency-dependent friction $\widetilde{\Gamma}(f)$  (blue solid line).}
\label{oh_friction}
\end{figure*}

\begin{figure*}[htp]
\centering
\begin{overpic}[width=0.49\textwidth]{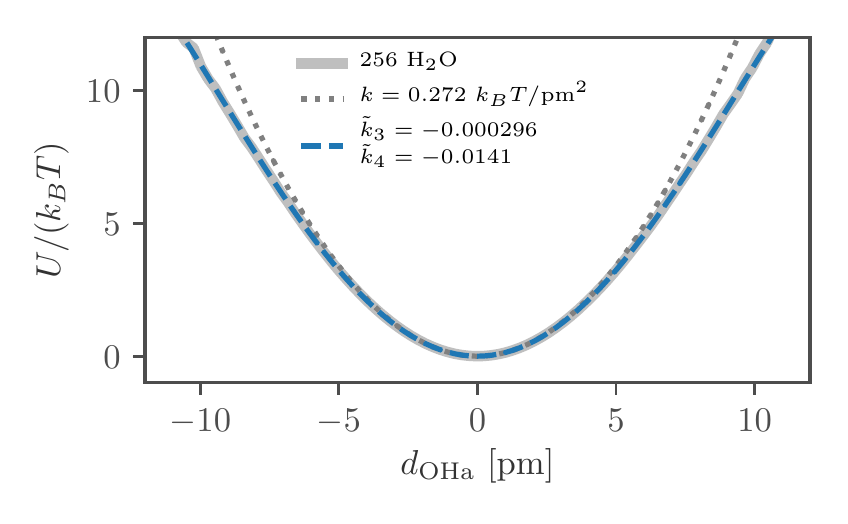}
\put(2,55){\large \bf A}
\end{overpic}
\begin{overpic}[width=0.49\textwidth]{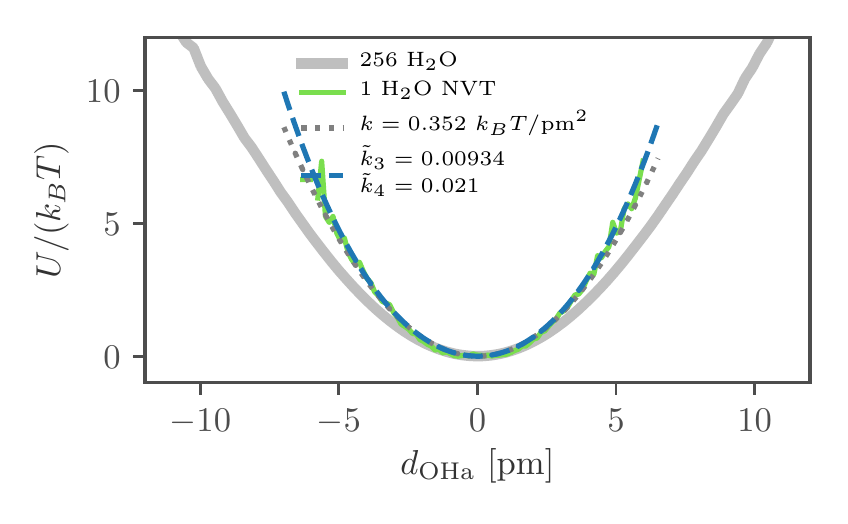}
\put(2,56){\large \bf B}
\end{overpic}
\begin{overpic}[width=0.49\textwidth]{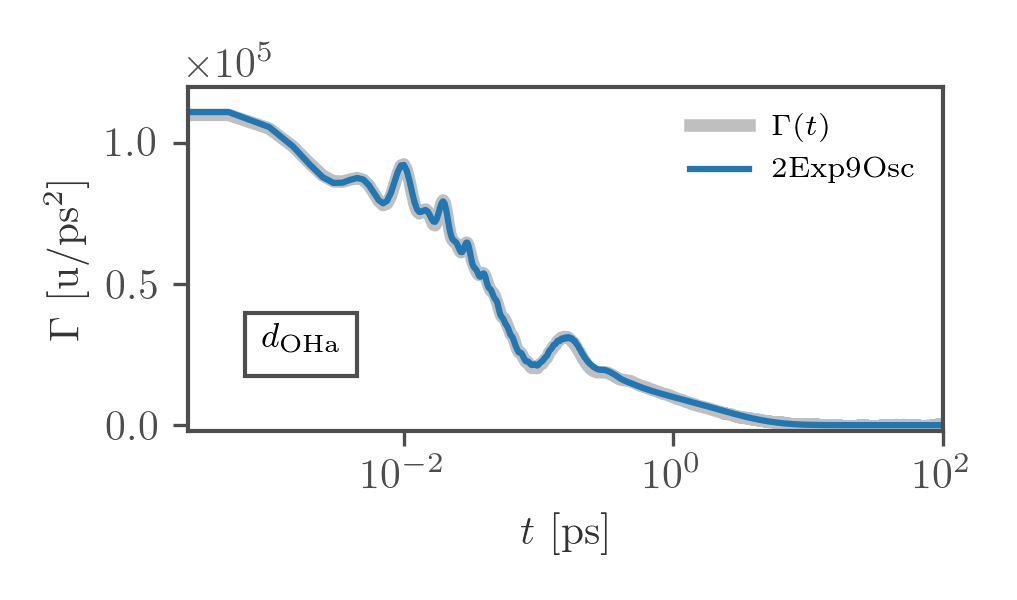}
\put(2,56){\large \bf C}
\end{overpic}
\begin{overpic}[width=0.49\textwidth]{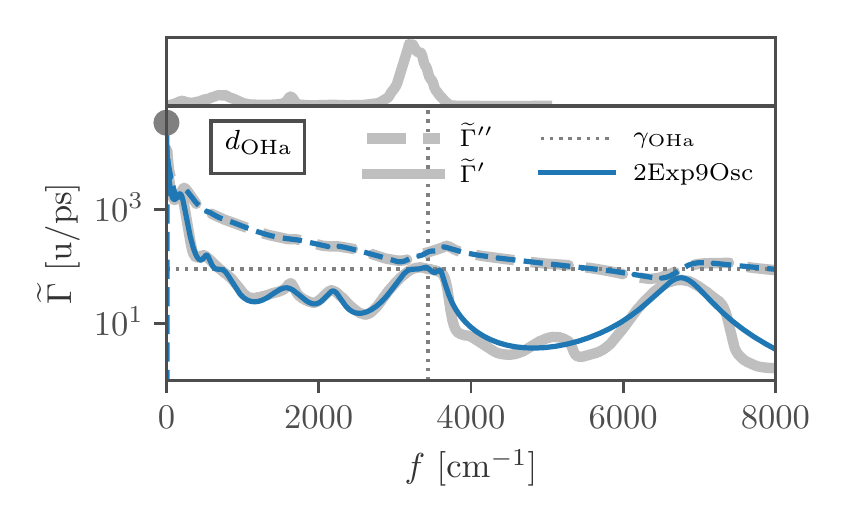}
\put(2,56){\large \bf D}
\end{overpic}
\begin{overpic}[width=0.49\textwidth]{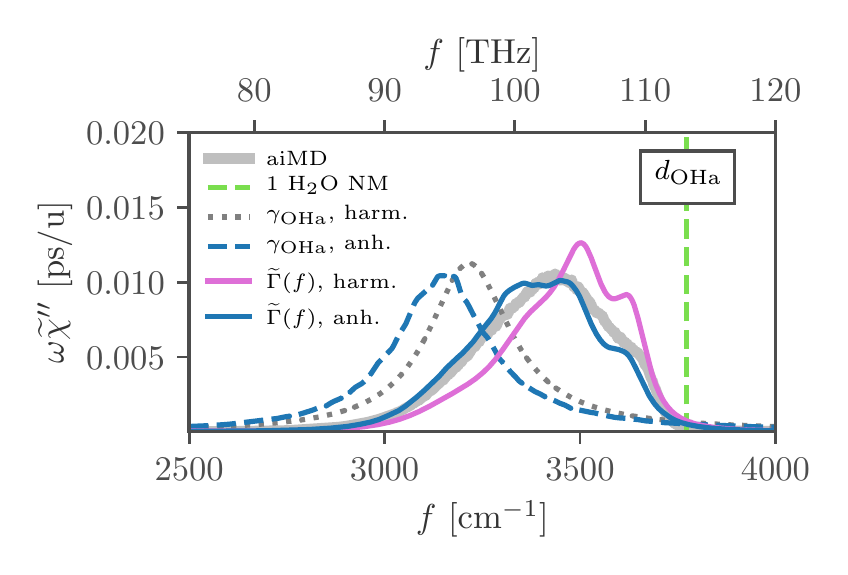}
\put(2,55){\large \bf E}
\end{overpic}
\caption{\footnotesize Results for the anti-symmetric stretch coordinate  $\doha$  from aiMD simulations.
{\bf A, B:} Potential $U(\doha)$ for 256 H$_2$O in the liquid phase (grey solid line) and for  a single H$_2$O  (green solid line in B), both
compared to the non-harmonic fit according to eq.~\eqref{eq:pot_param} in the main text (blue broken line) and the harmonic part  (grey dotted line).
{\bf C, D:} Friction as a function of time and  frequency  (grey lines) compared with
the fit according to eq.~\eqref{eq:kernel_param} in the main text (blue lines). Real and imaginary parts in (D) are shown as solid and broken lines,
the spectrum on top is the full absorption spectrum from aiMD. The dotted horizontal line in (D) shows the constant  real friction $\gamma_{\rm OHa}= \widetilde{\Gamma}'(f_{\rm OHa})$ evaluated at the OH stretch vibrational frequency $f_{\rm OHa}=\SI{3440}{cm^{-1}}$.
The grey circle denotes the static friction $\widetilde \Gamma'(0)$.
{\bf E:} Power spectrum $\omega \widetilde{\chi}''$
(grey solid line) compared to  models of varying complexity:
normal mode of single H$_2$O (broken vertical line),
Lorentzian with harmonic potential and constant friction $\gamma_{\rm OHa}$  (grey dotted line),
non-harmonic potential  and constant friction $\gamma_{\rm OHa}$  (blue broken line),
harmonic potential and frequency-dependent friction $\widetilde{\Gamma}(f)$ (purple solid line),
non-harmonic potential  and frequency-dependent friction $\widetilde{\Gamma}(f)$  (blue solid line).}
\label{oha_friction}
\end{figure*}

The potentials, power spectra and analysis in terms of frequency-dependent friction of the $\dohs$ and $\hoh$ coordinates are shown in the main text. The remaining discussion is shown here for the $\doh$ coordinate in fig.~\ref{oh_friction} and for the $\doha$ coordinate in fig.~\ref{oha_friction}. The potential of the $\doh$ coordinate shows strong non-harmonic contributions in fig.~\ref{oh_friction}A, seen from the reduced potential coefficients $\tilde k_3 = k_3/k_BT (k/k_BT)^{-3/2} = -0.128$ and  $\tilde k_4 = k_4/k_BT (k/k_BT)^{-2} = 0.00503$. The potential of the $\doha$ coordinate is symmetric by definition and therefore has a negligible cubic and only a small quartic contribution $\tilde k_4 = -0.0141$ in fig.~\ref{oha_friction}A, when comparing  to the potential of the $\dohs$ coordinate, shown in fig.~\ref{ohs_friction}A in the main text to have a significant cubic contribution.
When compared to the potential of a single water molecule in figs.~\ref{oh_friction}B and ~\ref{oha_friction}B, the $\doh$ coordinate shows a shift of the minimum from \SI{97.37}{pm} to \SI{99.06}{pm}, i.e. elongation of the bond length due to hydrogen bonding. As expected the potential of $\doha$ coordinate is centered around zero for both systems. Both coordinates show a significant potential softening in the liquid phase, which follows from comparison of the dominant harmonic contributions of $k/(k_BT)=\SI{0.139}{pm^{-2}}$ for $\doh$ and $k/(k_BT)=\SI{0.272}{pm^{-2}}$ for $\doha$ with the respective gas-phase values of $k/(k_BT)=\SI{0.183}{pm^{-2}}$ for $\doh$ and $k/(k_BT)=\SI{0.352}{pm^{-2}}$ for $\doha$.

The memory kernels in the time domain, shown in figs.~\ref{oh_friction}C for the $\doh$ coordinate and ~\ref{oha_friction}C for the $\doha$ coordinate, are discussed further below in fig.~\ref{oh_decomp_kernels}. The frequency-dependent friction, shown in figs.~\ref{ohs_friction}D in the main text for the $\dohs$ coordinate, ~\ref{oh_friction}D for the $\doh$ coordinate and ~\ref{oha_friction}D for the $\doha$ coordinate show remarkable differences. The frequency-dependent friction of the $\dohs$ coordinate shows a much stronger contribution in the regime of the HOH bending band at around \SI{1650}{cm^{-1}}, indicating that it couples more intensely to the HOH bending mode than the $\doha$ mode.

Figs.~\ref{oh_friction}E and \ref{oha_friction}E show the power spectra of the two coordinates, compared to the different \ac{GLE} models. As shown in the main text already for the $\dohs$ and $\hoh$ coordinates, the frequency-dependent-friction model including an non-harmonic potential outperforms the other models for all coordinates. In case of the $\doha$ coordinate, shown in fig.~\ref{oha_friction}E, the frequency-dependent-friction model with the harmonic potential also performs very well, due to the negligible non-harmonic contribution to the potential in fig.~\ref{oha_friction}A.

\begin{figure}
\begin{overpic}[width=0.49\textwidth]{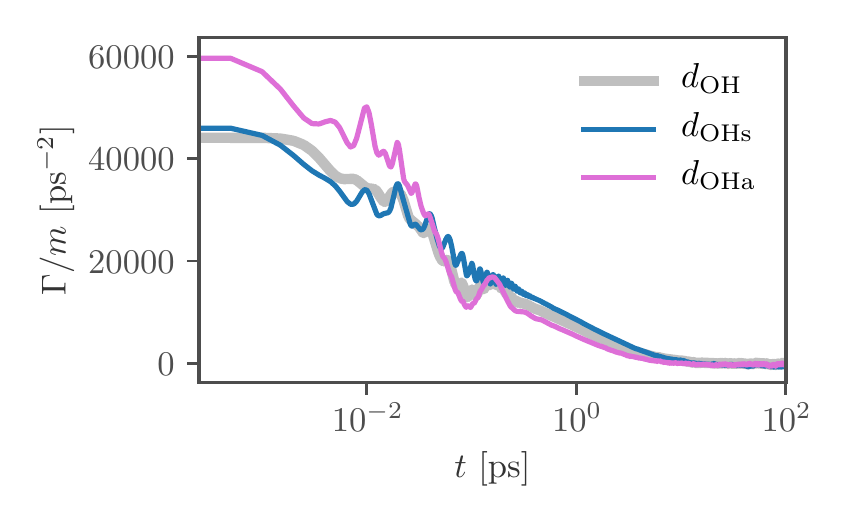}
\end{overpic}
\caption{Comparison of time-dependent friction kernels extracted from the \ac{aiMD} simulations for the $\doh$ (grey solid line), $\dohs$ (blue solid line) and $\doha$ (purple solid line) coordinates.}
\label{oh_decomp_kernels}
\end{figure}

The time-dependent friction kernels of the $\doh$, $\dohs$ and $\doha$ coordinates are compared in fig.~\ref{oh_decomp_kernels}, scaled by the mass of the respective coordinate. The time-dependent friction kernel of the $\doha$ coordinate differs significantly from the one of the $\dohs$ coordinate. However, all friction kernels show various oscillating decay time scales between \SI{10}{fs} and \SI{5}{ps}, which is discussed in detail the main text for the example of $\dohs$.

\section{Constant-friction line shape}
\label{deltaExampleSection}

\begin{figure*}[htb]
\centering
\begin{overpic}[width=0.49\textwidth]{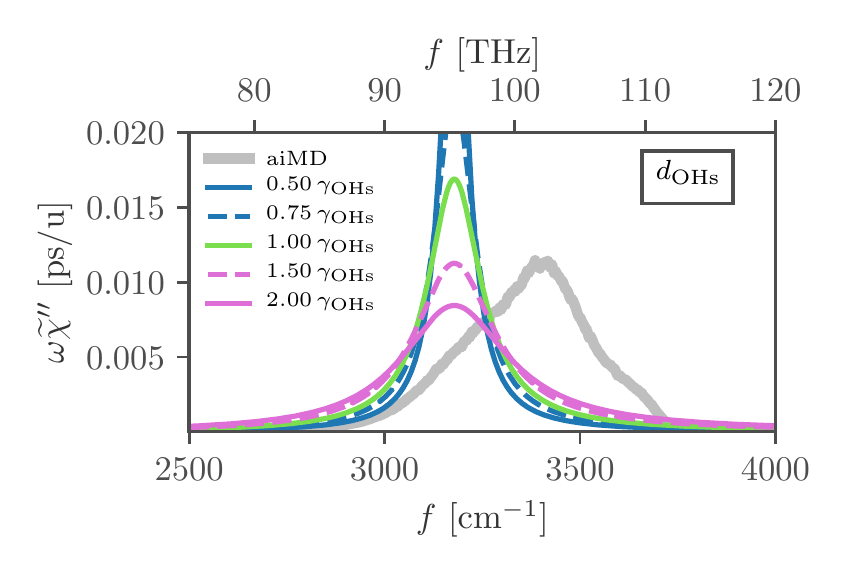}
\put(2,55){\large \bf A}
\end{overpic}
\begin{overpic}[width=0.49\textwidth]{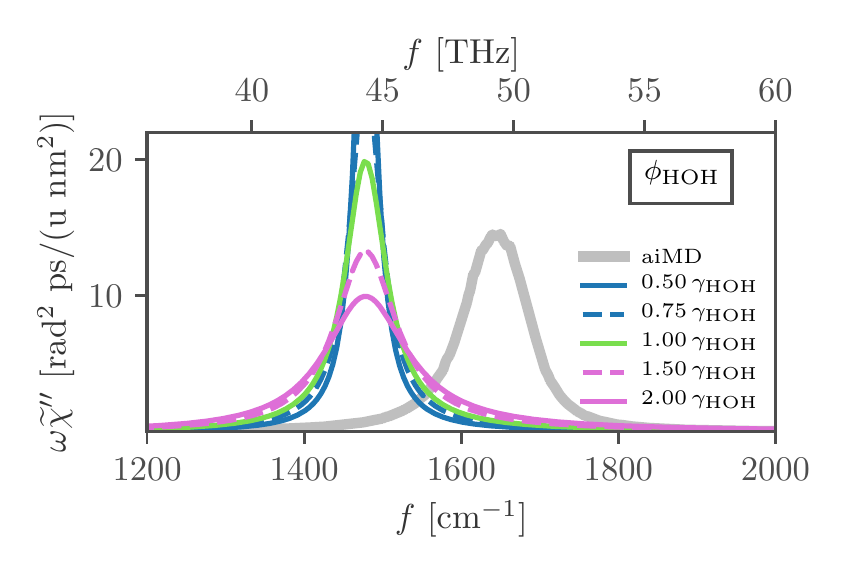}
\put(2,55){\large \bf B}
\end{overpic}
\begin{overpic}[width=0.49\textwidth]{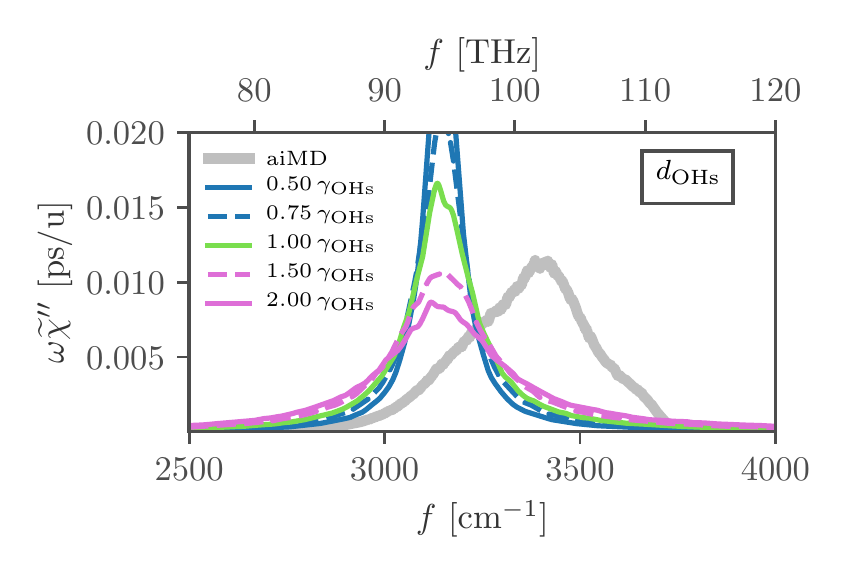}
\put(2,55){\large \bf C}
\end{overpic}
\begin{overpic}[width=0.49\textwidth]{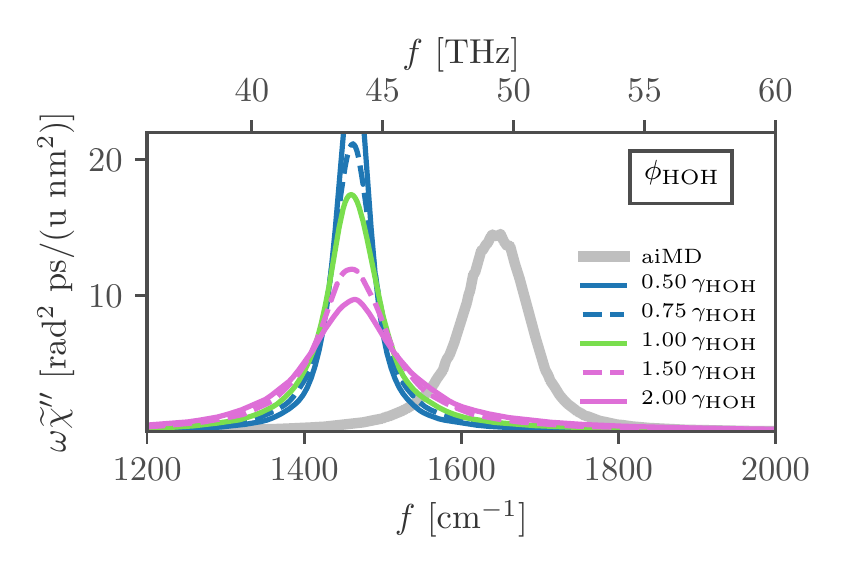}
\put(2,55){\large \bf D}
\end{overpic}
\caption{Power spectra, obtained from the \ac{aiMD} simulations (grey solid lines) and compared to spectra from a constant friction model with the harmonic fit to the potential (A, B) and considering non-harmonic effects (C, D). The spectra are shown for variations of the friction coefficient $\gamma_{\rm OHS} = \Gamma'(f_{\rm OHS})$ and $\gamma_{\rm HOH} = \Gamma'(f_{\rm OHS})$, where  $f_{\rm OHS}=\SI{3390}{cm^{-1}}$ and $f_{\rm HOH}=\SI{1650}{cm^{-1}}$ are at the maxima of the respective power spectra from the simulations. {\bf A, C:} $\dohs$ coordinate and {\bf B, D:} $\hoh$ coordinate.}
\label{delta_example}
\end{figure*}

The dependence of the Lorentzian line shape function eq.~\eqref{LinResHORed} on the friction coefficient $\gamma$ for the $\dohs$ coordinate is illustrated in fig.~\ref{delta_example}A and for the $\hoh$ coordinate in fig.~\ref{delta_example}B. A variation of $\gamma$ does not shift the peak position but significantly changes the width. This follows analytically from the maximum of eq.~\eqref{LinResHORed}
\begin{align}
\frac{\partial}{\partial \omega}\left[ \omega \widetilde \chi''(\omega) \right]  &= \frac{L^2}{k_BT}\left[-\frac{8 \tau  \omega  \left(\omega ^4 \tau _{\omega }^4-1\right)}{\left(\omega ^2 \left(\tau ^2-8 \tau _{\omega }^2\right)+4 \omega ^4 \tau _{\omega }^4+4\right){}^2}\right] \\
 &\overset{!}{=} 0\quad \text{for}\quad \omega=\tau_{\omega}^{-1},
\end{align}
which does not depend on $\gamma$. When considering non-harmonic effects in the potential, simulated power spectra still show no significant shifts for variation of the friction coefficient, as shown in figs.~\ref{delta_example}C and D.

\clearpage

\section{Simulations of single H$_2$O in the NVE ensemble}
\label{initDistSection}

\begin{figure*}[htb]
\centering
\begin{overpic}[width=0.49\textwidth]{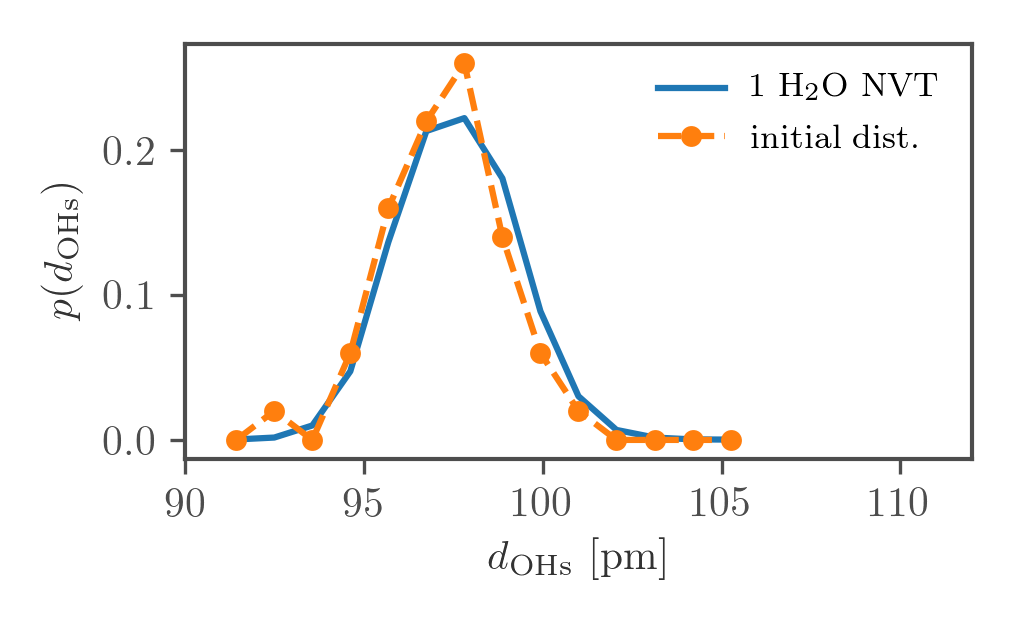}
\put(2,55){\large \bf A}
\end{overpic}
\begin{overpic}[width=0.49\textwidth]{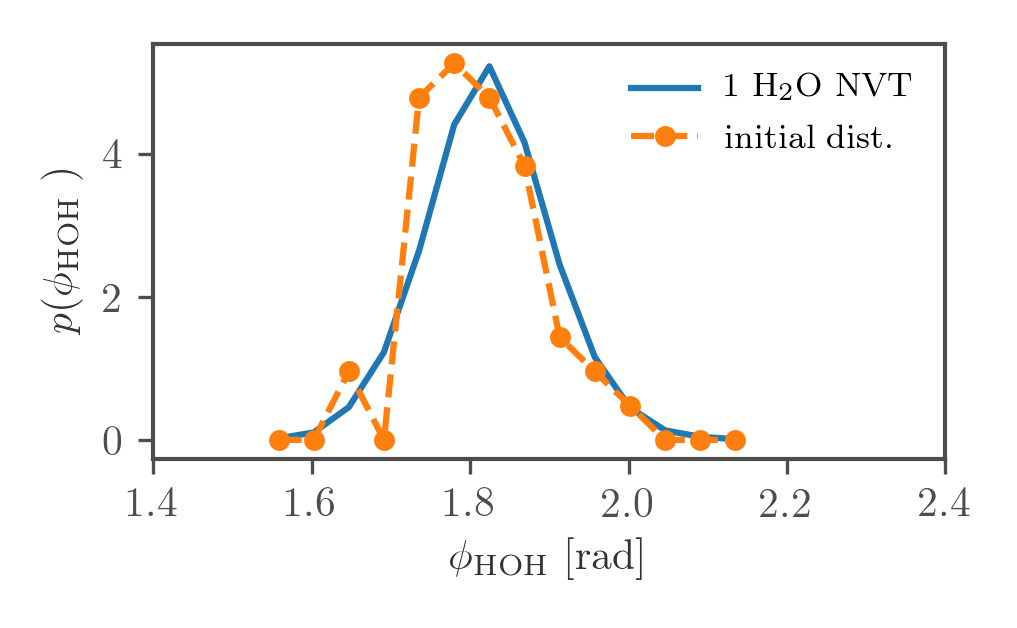}
\put(2,55){\large \bf B}
\end{overpic}
\begin{overpic}[width=0.49\textwidth]{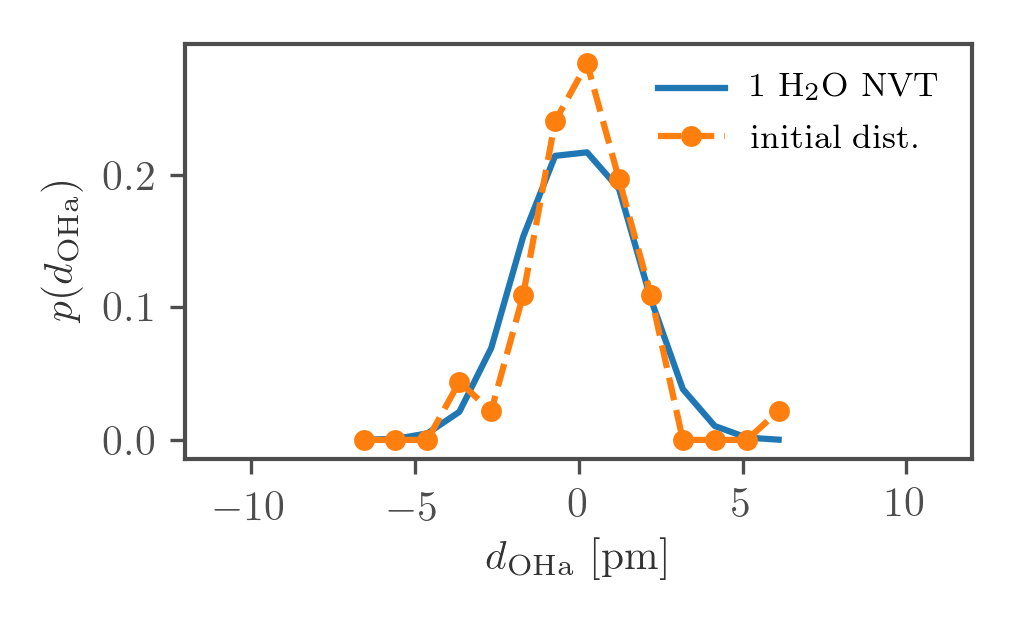}
\put(2,55){\large \bf C}
\end{overpic}
\begin{overpic}[width=0.49\textwidth]{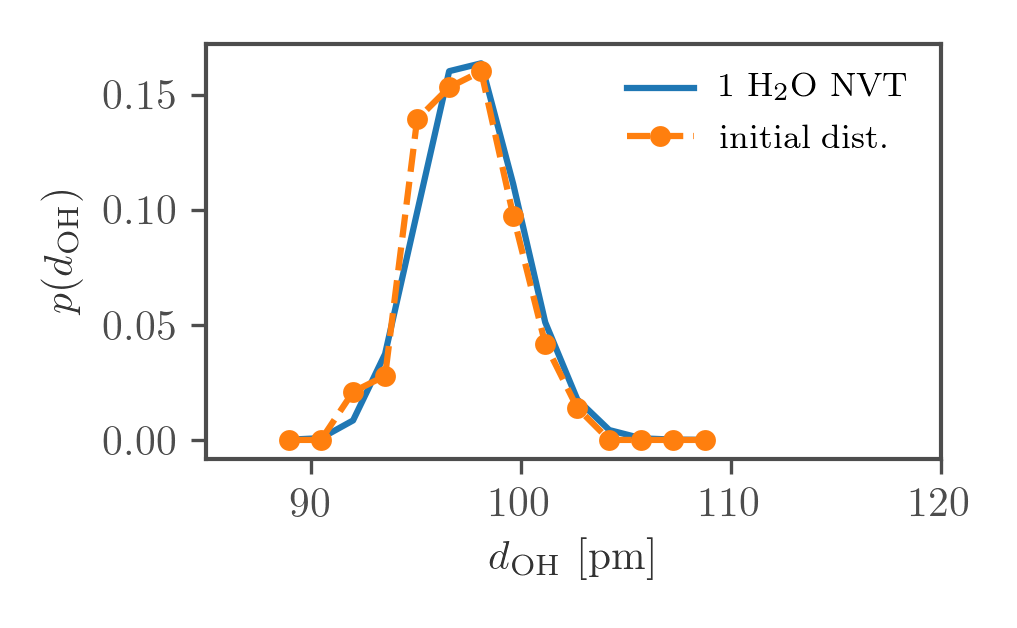}
\put(2,55){\large \bf D}
\end{overpic}
\caption{Comparison of the distributions for the different coordinates sampled from the \ac{aiMD} simulations of a single H$_2$O molecule under NVT conditions (blue solid lines) and 47 samples taken from the NVT trajectory as initial conditions for equivalent simulations under NVE conditions (orange dots). {\bf A:} $\dohs$ coordinate, {\bf B:} $\hoh$ coordinate, {\bf C:} $\doha$ coordinate and {\bf D:} $\doh$ coordinate.}
\label{1nve_dist}
\end{figure*}

For the \ac{aiMD} simulations of single H$_2$O molecules, representing the gas phase, 47 initial configurations were sampled from a \SI{25}{ps} NVT simulation and subsequently simulated in the NVE ensemble. The NVT simulation was temperature-controlled using an individual thermostat with a time constant of \SI{10}{fs} for each atom. The NVE simulations were each run for \SI{10}{ps} with a time step of \SI{0.25}{fs}. The distributions of their initial configurations are shown in fig.~\ref{1nve_dist} to sample well the equilibrium distributions from the NVT trajectory.

\clearpage

\section{Wiener-Khintchine theorem}
\label{WienerKintchineSection}

The correlation function $C_{xy}(t)$ of two stochastic processes $x(t)$ and $y(t)$ limited to the interval $[0,L_t]$ is efficiently computed from the Fourier-transformed expressions $\tilde x(\omega)$ and $\tilde y(\omega)$ according to
\begin{align}
\label{WienerKintchine}
C_{xy}(t) = \frac{1}{2 \pi (L_t-t)} \int_{-\infty}^{\infty} d\omega\ e^{-i\omega t} \tilde x(\omega) \tilde y^*(\omega),
\end{align}
where the asterisk denotes the conjugate form. This is known as the Wiener-Khintchine theorem \cite{Wiener1930}. Both sides of eq.~\eqref{WienerKintchine} are Fourier-transformed to give
\begin{align}
\int_{-\infty}^{\infty} dt\ e^{i\omega t}\ 2 \pi L_t \left( 1-\frac{t}{L_t}\right) C_{xy}(t) &= \tilde x(\omega) \tilde y^*(\omega),
\end{align}
which in the limit of large $L_t$ reduces to
\begin{align}
\label{WienerKintchineFT}
\widetilde C_{xy}(\omega) &= L_t^{-1} \tilde x(\omega) \tilde y^*(\omega).
\end{align}
Eq.~\eqref{WienerKintchine} can be derived starting off with the definition of the correlation function
\begin{align}
C_{xy}(t) = \frac{1}{L_t-t}\int_{0}^{L_t-t} dt'\ x(t'+t) y(t'),
\end{align}
and making use of the convolution theorem
\begin{align}
C_{xy}(t) &= \frac{1}{4 \pi^2 (L_t-t)} \int_{0}^{L_t-t} dt' \nonumber \\
 &\ \ \int_{-\infty}^{\infty} d\omega\ e^{-i\omega (t+t')} \tilde x(\omega) \int_{-\infty}^{\infty} d\omega'\ e^{-i\omega' t'} \tilde y(\omega') \nonumber \\
 &= \frac{1}{4 \pi^2 (L_t-t)} \int_{-\infty}^{\infty} d\omega\ e^{-i\omega t} \tilde x(\omega) \int_{-\infty}^{\infty} d\omega'\ \tilde y(\omega') \nonumber \\
 &\ \ \int_{0}^{L_t-t} dt'  e^{-it' (\omega+\omega')} \nonumber \\
 &= \frac{1}{4 \pi^2 (L_t-t)} \int_{-\infty}^{\infty} d\omega\ e^{-i\omega t} \tilde x(\omega) \int_{-\infty}^{\infty} d\omega'\ \tilde y(\omega') \nonumber \\
 &\ \ 2 \pi \delta(\omega+\omega') \nonumber \\
 &= \frac{1}{2 \pi (L_t-t)} \int_{-\infty}^{\infty} d\omega\ e^{-i\omega t} \tilde x(\omega) \tilde y(-\omega),
\end{align}
noting that $\tilde y(-\omega)=\tilde y^*(\omega)$ for a real function $y(t)$ in order to obtain eq.~\eqref{WienerKintchine}.

\clearpage
\bibliography{bibliography.bib}